\documentclass[12pt]{article}
\RequirePackage{ifpdf}
\ifpdf\RequirePackage[raiselinks=false,colorlinks=true,citecolor=blue,urlcolor=blue,linkcolor=blue,bookmarksopen=true,pdftex]{hyperref}\else
\RequirePackage[raiselinks=false,colorlinks=true,citecolor=blue,urlcolor=blue,linkcolor=blue,bookmarksopen=true,dvips]{hyperref}\fi
\usepackage[latin1]{inputenc} 
\usepackage[T1]{fontenc} 
\usepackage{amsfonts}
\usepackage{latexsym}
\usepackage{color}  %
\usepackage{ae,aecompl,amsbsy,amssymb}
\usepackage{epsf}
\usepackage{graphics}
\usepackage{framed}
\newcommand{\textctilde}{\raisebox{0.5ex}{\texttildelow}}

\newcommand{\limp}{\Rightarrow}
\newcommand{\Coq}{{\sc Coq}}
\newcommand{\Ssreflect}{{\sc SSReflect}}
\usepackage{tikz}
\usetikzlibrary{arrows.meta,calc,decorations.markings,math,arrows.meta}
\newlength{\hsbw}
\newcommand\MSpacing{13pt}

\newenvironment{boxed}{\begin{flushleft}
\setlength{\hsbw}{\textwidth}
\addtolength{\hsbw}{-\arrayrulewidth}
\addtolength{\hsbw}{-\tabcolsep}
 \begin{tabular}{@{}|c@{}|@{}}\hline 
 \begin{minipage}[b]{\hsbw}
 \vspace*{.06in}
 \begingroup\small\baselineskip\MSpacing}{\endgroup\end{minipage}\\ \hline 
 \end{tabular}
 \end{flushleft}}

\newcommand{\arrowIn}{
\tikz \draw[-{Stealth[scale=1.2]}] (-1pt,0) -- (1pt,0);
}

\title{A Formalisation of Algorithms for Sorting Network}
\author{Laurent Th{\'e}ry \\
{\tt Laurent.Thery@sophia.inria.fr}}
\date{}
\begin{document}
\maketitle
\begin{abstract}

This notes explains how standard algorithms that construct 
sorting networks have been formalised and proved correct in  the 
{\Coq} proof assistant using the {\Ssreflect} extension.
\end{abstract}

\section{Introduction}

A network is composed of a number of lines. By analogy to electronic
circuit, each line has an input value before entering the network
and an output value when leaving the network. The building block of a network
is a comparator. A comparator connects two lines
\vskip5pt
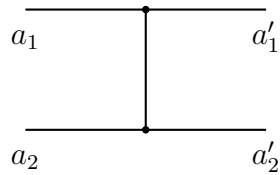
\begin{center}
\begin{tikzpicture}[scale=0.80]
\foreach \a in {1,2}
  \draw[thick] (0,5 - 2 * \a) node[anchor=north] {\vphantom{$a'$}$a_\a$} -- ++(4,0) 
  node[anchor=north] {$a'_\a$};;
\foreach \x in {{2,1},{2,3}}
  \filldraw (\x) circle (1.5pt);
\draw[thick] (2,1) -- (2,3);

\end{tikzpicture}
\end{center}
\vskip5pt
\noindent
A connector works as follows.
The output value of the upper line is the minimun of the input values,
$a'_1 = \min(a_1, a_2)$.
The output value of the lower line is the maximum of the two lines,
$a'_1 = \max(a_1, a_2)$.

A network is a collection of connectors. Here, we are interested
into networks that sort their inputs, i.e. they return sorted outputs. 
An example of a network that sorts 3 inputs is the following
\vskip5pt
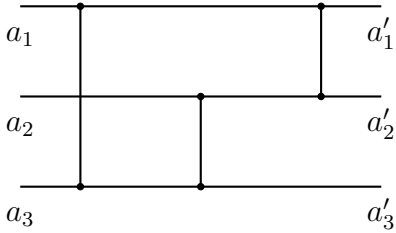
\begin{center}
\begin{tikzpicture}[scale=0.80]
\foreach \a in {1,2,3}
  \draw[thick] (0, 5.5 - 1.5 * \a) node[anchor=north] {\vphantom{$a'$}$a_\a$} -- ++(6,0) 
  node[anchor=north] {$a'_\a$};;
\foreach \x in {{1,1},{1,4}, {3,1},{3,2.5}, {5,2.5},{5,4}}
  \filldraw (\x) circle (1.5pt);
\draw[thick] (1,1) -- (1,4);
\draw[thick] (3,1) -- (3,2.5);
\draw[thick] (5,2.5) -- (5,4);
\end{tikzpicture}
\end{center}
\vskip5pt
\noindent
Whatever the initial values $a_1$, $a_2$ and $a_3$ are, we have
$a'_1 \le a'_2 \le a_3$. In the rest of the paper we are interesting
in proving the correctness of some recursive algorithms that build 
sorting network. 
We first explain how we have formalised networks. Then, we present
3 algorithms:
\begin{itemize}
\item[-] an algorithm that builds the bitonic sorting network;  
\item[-] an algorithm that builds the odd-even merge sorting network;  
\item[-] an algorithm that builds the odd-even exchange sorting network.  
\end{itemize}

\section{The formalisation}

In the formalisation, we are using material that comes 
from the Mathematical Component Library. In order to make the presentation
understandable by someone not familiar with this library, we summarize
in the appendices \ref{basic}, \ref{fintype}, \ref{seq} and \ref{tuple}
the notions that have been used for this formalisation.

To represent the state of lines, we are using the \texttt{tuple} type
and are working on an arbitrary \texttt{orderedType} $A$. So if the 
network has $m$ lines, the state of lines is represented by 
a $m$\texttt{.-tuple}\, $A$. We allow connectors to work simultaneously on 
several disjoint pairs of lines. If we consider the following sequence composed 
of 3 connectors
\vskip5pt
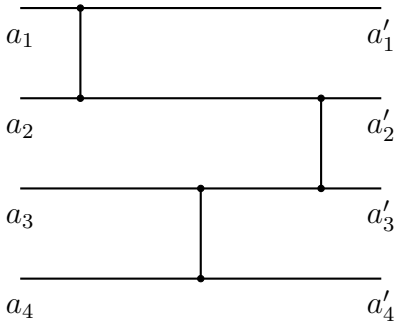
\begin{center}
\begin{tikzpicture}[scale=0.80]
\foreach \a in {1,2,3,4}
  \draw[thick] (0, 7 - 1.5 * \a) node[anchor=north] {\vphantom{$a'$}$a_\a$} -- ++(6,0) 
  node[anchor=north] {$a'_\a$};;
\foreach \x in {{1,5.5},{1,4}, {3,2.5},{3,1}, {5,4},{5,2.5}}
  \filldraw (\x) circle (1.5pt);
\draw[thick] (1,5.5) -- (1,4);
\draw[thick] (3,1) -- (3,2.5);
\draw[thick] (5,2.5) -- (5,4);
\end{tikzpicture}
\end{center}
\vskip5pt
\noindent
the first two are independent so can be performed in parallel while
the third one must be kept separated as it shares some lines with 
the previous two. Making the parallelism explicit, we get the following 
drawing with only 2 connectors.
\vskip5pt
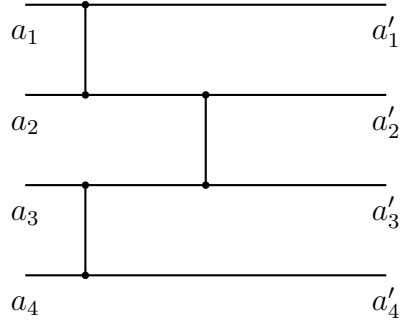
\begin{center}
\begin{tikzpicture}[scale=0.80]
\foreach \a in {1,2,3,4}
  \draw[thick] (0, 7 - 1.5 * \a) node[anchor=north] {\vphantom{$a'$}$a_\a$} -- ++(6,0) 
  node[anchor=north] {$a'_\a$};;
\foreach \x in {{1,5.5},{1,4}, {1,2.5},{1,1}, {3,4},{3,2.5}}
  \filldraw (\x) circle (1.5pt);
\draw[thick] (1,5.5) -- (1,4);
\draw[thick] (1,1) -- (1,2.5);
\draw[thick] (3,2.5) -- (3,4);
\end{tikzpicture}
\end{center}
\vskip5pt
\noindent
A connector is then encoded as a record that contains a function \textit{clink}
that takes a line (an element of \texttt{I\_}$m$) 
and returns its associated line. The function is the identity for lines
that are not connected.
The requirement of the lines to be associated in disjoint pairs is encoded
in the \textit{cfinv} field which asks for \textit{clink} to be involutive. 
A network is then a list of connectors.
\vskip0pt
\noindent
\begin{framed}
\footnotesize
\noindent
\texttt{Record} \textit{connector} ($m$ : \texttt{nat}) := 
  \textit{connector\_of} \{ 
\vskip0pt\noindent\hskip10pt
\textit{clink} \texttt{:} \{\texttt{ffun} \texttt{I\_}$m$ 
  $\limp$ \texttt{I\_}$m$\}\texttt{;}
\vskip0pt\noindent\hskip10pt
  \textit{cfinv} : \texttt{[forall} $i$\texttt{,} 
  \texttt{clink} (\texttt{clink} $i$\texttt{) ==} $i$\texttt{]}\\ 
\}.
\vskip5pt\noindent\hskip0pt
\texttt{Definition} \textit{network} \texttt{:=} \texttt{seq} (\textit{connector} $m$).
\end{framed}
\vskip0pt
\noindent
An example of such a connector is the one that swaps the value of two 
line $i$ and $j$. Its definition is done in three steps. We first 
define the link function, we prove that it is involutive, and we finally
build the connector.
\vskip0pt
\noindent
\begin{framed}
\footnotesize
\noindent
\vskip0pt\noindent\hskip0pt
\texttt{Definition} \textit{clink\_swap} ($i$ $j$\texttt{:} \texttt{I\_}$m$) 
  \texttt{:} \texttt{\{ffun} \texttt{I\_}$m$ \texttt{->} 
  \texttt{I\_}$m$\texttt{\}} \texttt{:=}
\vskip0pt\noindent\hskip10pt
  \texttt{[ffun} $x$ \texttt{=>} \texttt{if} $x$ \texttt{==} $i$ 
  \texttt{then} $j$ \texttt{else} \texttt{if} $x$ \texttt{==} $j$ 
  \texttt{then} $i$ \texttt{else} $x$\texttt{]}. 
\vskip0pt\noindent\hskip0pt
\texttt{Lemma} \texttt{clink\_swap\_proof} ($i$ $j$ \texttt{:} 
  \texttt{I\_}$m$) : 
\vskip0pt\noindent\hskip10pt
  \texttt{[forall} $k$, \textit{clink\_swap} $i$ $j$ 
  (\textit{clink\_swap} $i$ $j$ $k$) \texttt{==} $k$\texttt{]}.
\vskip0pt\noindent\hskip0pt
\texttt{Definition} \textit{cswap} $i$ $j$ \texttt{:=}
\textit{connector\_of} (\textit{clink\_swap\_proof} $i$ $j$).
\end{framed}
\vskip0pt
\noindent

In the following, a variable $c$ always represents a connector, $n$ a network,
$s$ a sequence and $t$ 
a tuple. The first operation on connector and network is the one that 
computes output values.
The function \textit{cfun} applies a connector $c$ to a tuple $t$ 
and the function \textit{nfun} applies a network $n$ to a tuple $t$.
\vskip0pt
\noindent
\begin{framed}
\footnotesize
\noindent
\vskip0pt\noindent\hskip0pt
\texttt{Definition} \textit{cfun} $c$ $t$ \texttt{:=}
\vskip0pt\noindent\hskip10pt
    \texttt{[tuple if } $i$ $\le$ \texttt{clink} $c$ $i$
\vskip0pt\noindent\hskip50pt
           \texttt{then} \texttt{min} (\texttt{tnth} $t$ $i$) 
           (\texttt{tnth} $t$ (\textit{clink} $c$ $i$))
\vskip0pt\noindent\hskip50pt
           \texttt{else} \texttt{max} (\texttt{tnth} $t$ $i$)
              (\texttt{tnth} $t$ (\textit{clink} $c$ $i$)) 
           \texttt{|} $i$ \texttt{<} $m$\texttt{]}.
\vskip5pt\noindent\hskip0pt
\texttt{Definition} \textit{nfun} $n$ $t$ 
\texttt{:=} \texttt{foldl} 
(\texttt{fun} $t$ $c$ \texttt{=>} \textit{cfun} $c$ $t$) $t$ $n$.
\end{framed}
\vskip0pt
\noindent
The function \textit{cfun} performs the swap between values of connected
lines, while \textit{nfun} simply iterates the application of \textit{cfun}.

The first obvious property of connector and network is that they only
permute their outputs. This is proved by the following theorems
\vskip0pt
\noindent
\begin{framed}
\footnotesize
\noindent
\vskip0pt\noindent\hskip0pt
\texttt{Lemma} \textit{perm\_cfun} $c$ $t$ \texttt{:} 
\texttt{perm\_eq} (\textit{cfun} $c$ $t$) $t$.
\vskip0pt\noindent\hskip0pt
\texttt{Lemma} \textit{perm\_nfun} $n$ $t$ \texttt{:} 
\texttt{perm\_eq} (\textit{nfun} $n$ $t$) $t$.
\end{framed}
\vskip0pt
\noindent
Another interesting property is the regularity with respect to the order.
If we take two arbitrary ordered types $A$ and $B$ and $f$ a function 
from $A$ to $B$ that behaves well with the order ($f\,\, x \le_{B} f\,\, y$ iff $x \le_{A} y$)
we have the following properties for \texttt{min} and \texttt{max} :
\noindent
\vskip0pt
\noindent
\begin{framed}
\footnotesize
\noindent
\vskip0pt\noindent\hskip0pt
\texttt{Lemma} \textit{min\_homo} ($x$ $y$ \texttt{:} $A$) \texttt{:} $f$ (\texttt{min} $x$ $y$) =
\texttt{min} ($f$ $x$) ($f$ $y$).
\vskip0pt\noindent\hskip0pt
\texttt{Lemma} \textit{max\_homo} ($x$ $y$ \texttt{:} $A$)  \texttt{:} $f$ (\texttt{max} $x$ $y$) =
\texttt{min} ($f$ $x$) ($f$ $y$).
\end{framed}
\vskip0pt
\noindent
These properties can then be easily lifted at the level of connector and network.
\noindent
\begin{framed}
\footnotesize
\noindent
\vskip0pt\noindent\hskip0pt
\texttt{Definition} \textit{tmap} $f$ $t$ \texttt{:=} 
\texttt{[tuple} $f$ (\texttt{tnth} $t$ $i$) \texttt{|} $i$ \texttt{<} $m$\texttt{]}.
\vskip0pt\noindent\hskip0pt
\texttt{Lemma} \textit{tmap\_connector} $c$ ($t$ : $m$\texttt{.-tuple} $A$) : 
   \textit{tmap} $f$ (\textit{cfun} $c$ $t$) = 
   \textit{cfun} $c$ (\textit{tmap} $f$ $t$).
\vskip0pt\noindent\hskip0pt
\texttt{Lemma} \textit{tmap\_network} $n$ ($t$ : $m$\texttt{.-tuple} $A$) : 
  \textit{tmap} $f$ (\textit{nfun} $n$ $t$) =
  \textit{nfun} $n$ (\textit{tmap} $f$ $t$).
\end{framed}
\vskip0pt
We are now ready to define the notion of sorting network.
It is defined as a qualifier so we express the fact the $n$ is a 
sorting network by the expression "$n$ \texttt{is} \textit{sorting}".
Thanks to the regularity with respect to the order, we can limit the
definition of being a sorting network to the one of sorting all the
boolean tuples. As, if we consider $m$ lines,
there are only a finite number of such tuples ($2^m$ to be precise), 
this property is decidable and can be encoded as a boolean.
\noindent
\begin{framed}
\footnotesize
\noindent
\vskip0pt\noindent\hskip0pt
\texttt{Definition} \textit{sorting} \texttt{:=}
\vskip0pt\noindent\hskip10pt
  \texttt{[qualify} $n$ \texttt{|} \texttt{[forall} $r$ \texttt{:}
   $m$\texttt{.-tuple} \texttt{bool}, \texttt{sorted}$_\le$ (\texttt{nfun} $n$ $r$)
   \texttt{]]}.
\end{framed}
\vskip0pt
We now need to show that this encoding covers exactly the usual notion
of sorting network. If we consider an arbitrary ordered type $A$, 
a network is sorting if and only if it sorts all the tuples of elements of $A$.
This is known as the zero-one principle.
One direction is straightforward. If there is at least two elements in $A$
sorting all the tuples in $A$ implies our definition.
\noindent
\begin{framed}
\footnotesize
\noindent
\vskip0pt\noindent\hskip0pt
\texttt{Lemma} \textit{sorted\_sorting} $n$ ($x_1$ $x_2$ : $A$) : 
\vskip0pt\noindent\hskip10pt
  $x_1$ \texttt{!=} $x_2$ $\limp$ ($\forall$$t$ \texttt{:}
   $m$\texttt{.-tuple} $A$, \texttt{sorted} $\le_A$ (\texttt{nfun} $n$ $t$)) 
   $\limp$ $n$ \texttt{is} \textit{sorting}.
\end{framed}
\vskip0pt\noindent
Given a boolean tuple $t$, if we consider the function $f$ from boolean to $A$
that returns \texttt{min} $x_1$ $x_2$ on \texttt{false} and 
\texttt{max} $x_1$ $x_2$ on \texttt{true}. Applying $f$ on the tuple $t_1$ gives us a tuple $t_1$
of elements of $A$. If we apply $n$ of $t_1$, it returns a sorted tuple 
$t_2$. Now, if we consider $g$ from $A$ to \texttt{bool} defined as 
$g\,\, x = \texttt{false}$ if $x \le \texttt{min}\,\, x_1\,\, x_2$ and 
\texttt{true} otherwise. It is easy to show that $g$ behaves well with the 
orders and is the left 
inverse of $f$ (we have $g\,\, (f\,\, b) = b$), so $\textit{tmap}\,\, g\,\, t_2$ is 
the result of applying the network $n$ to $t$ and is sorted.

Conversely, we have to reason by contradiction.
\noindent
\begin{framed}
\footnotesize
\noindent
\vskip0pt\noindent\hskip0pt
\texttt{Lemma} \textit{sorting\_sorted} $n$ ($t$ : $m$\texttt{.-tuple} $A$) :
  $n$ \texttt{is} \textit{sorting}
  $\limp$ 
  \texttt{sorted} $\le_A$ (\texttt{nfun} $n$ $t$)).
\end{framed}
\vskip0pt\noindent
Let us take an arbitrary tuple $t$ of elements of $A$.
Applying the network $n$ on $t$ gives a tuple $t_1$.
Suppose that $t_1$ is not sorted.
This means that there exists an $i$ such that $t_1[i] > t_1[i + 1]$.
If we consider $h$ from $A$ to \texttt{bool} that returns \texttt{false}
to elements strictly smaller than $t_1[i]$ and \texttt{true} otherwise.
Again, $h$ behaves well with the orders. So, 
\textit{tmap} $h$ $t$ is a boolean tuple $t$ whose application to $n$
gives \textit{tmap} $h$ $t_1$ which is not sorted by construction. 
This is in contradiction
with our assumpition of $n$ being a sorting network, so $t_1$ must be sorted.

Now, we are ready to build sorting networks. We first need building
blocks. A key block is the one that glues together two networks:
given a network $n_1$ with $m_1$ lines
and a network $n_2$ with $m_2$ lines, it creates a network with
 $m_1 + m_2$ lines that behaves like $n_1$ on the top lines and $n_2$
 on the bottom lines. There are different ways to do this. We favour 
 the one that tries to fuze together connectors. This is the one
 that will be handy for building our sorting network later.
So, at connector level, we have a connector $c_1$ with $m_1$
lines and a connector $c_2$ with $m_2$ lines and we want to 
build a connector of $m_1 + m_2$ lines. The first step is to 
build the associated \textit{clink}. This requires some surgery with ordinals.
Then, we need to prove that this new \textit{clink} is involutive and we
finally get our \textit{cmerge} operation.
\noindent
\begin{framed}
\footnotesize
\noindent
\vskip0pt\noindent\hskip0pt
\texttt{Definition} \textit{clink\_merge} $m_1$ $m_2$ 
($c_1$ : \textit{connector} $m_1$) ($c_2$ : \textit{connector} $m_2$) 
\texttt{:=}
\vskip0pt\noindent\hskip10pt
\texttt{[ffun} $i$ \texttt{=> match split} $i$ \texttt{with}
\vskip0pt\noindent\hskip60pt
             \texttt{| inl} $x$ \texttt{=>} \texttt{lshift} \texttt{\_} 
             (\texttt{clink} $c_1$ $x$)
\vskip0pt\noindent\hskip60pt
             \texttt{| inr} $x$ \texttt{=>} \texttt{rshift} \texttt{\_}
              (\texttt{clink} $c_2$ $x$)
\vskip0pt\noindent\hskip60pt
             \texttt{end]}.
\vskip5pt\noindent\hskip0pt
\texttt{Lemma} \textit{clink\_merge\_proof} $m_1$ $m_2$ ($c_1$ : 
\textit{connector} $m_1$) ($c_2$ : \textit{connector} $m_2$) :
\vskip0pt\noindent\hskip10pt
  \texttt{[forall} $i$, (\textit{clink\_merge} $c_1$ $c_2$ 
  (\textit{clink\_merge} $c_1$ $c_2$ $i$)) \texttt{==} $i$\texttt{]}.
\vskip5pt\noindent\hskip0pt
\texttt{Definition} \textit{cmerge} $m_1$ $m_2$ ($c_1$ : 
\textit{connector} $m_1$) ($c_2$ : \textit{connector} $m_2$) \texttt{:=} 
\vskip0pt\noindent\hskip10pt
  \textit{connector\_of} (\textit{clink\_merge\_proof} $c_1$ $c_2$).
\end{framed}
\vskip0pt\noindent
Lifting this to network is easier. 
We create the sequence of pairs of connectors of $n_1$  and $n_2$ and
on each of these pairs we apply \textit{cmerge}.
\noindent
\begin{framed}
\footnotesize
\noindent
\vskip0pt\noindent\hskip0pt
\texttt{Definition} \texttt{nmerge} $m_1$ $m_2$ 
($n_1$ \texttt{:} \textit{network} $m_1$) ($n_2$ : 
\textit{network} $m_2$) \texttt{:=}
\vskip0pt\noindent\hskip10pt
    \texttt{[seq} \textit{cmerge} i\texttt{.1} i\texttt{.2}
    \texttt{|} $i$ \texttt{<-} \texttt{zip} $n_1$ $n_2$\texttt{]}.
\end{framed}
\vskip0pt\noindent
Note that this construction really makes sense of $n_1$ and $n_2$ have the 
same numbers of connectors. Otherwise the \texttt{zip} operation looses some 
connectors of the longest network. As a matter of fact, in the following, we
mostly use the duplication operator that glues together two identical pieces.
\noindent
\begin{framed}
\footnotesize
\noindent
\vskip0pt\noindent\hskip0pt
\texttt{Definition} \texttt{ndup} $m$ ($n$ \texttt{:} \textit{network} $m$) : 
\textit{network} ($m + m$) \texttt{:=} \textit{cmerge} $n$ $n$
\end{framed}
\vskip0pt\noindent
Another way of gluing network is the one based on parity. Given
$n_1$ and $n_2$, we build a network $n$ whose even lines are ruled by $n_1$ and
the odd ones by $n_2$. We first need to introduce the division by 2 
and the even and odd doubling at the level of ordinals.
\noindent
\begin{framed}
\footnotesize
\noindent
\vskip0pt\noindent\hskip0pt
\texttt{Definition} \textit{idiv2} $m$ \texttt{:} {\textquotesingle}\texttt{I\_}($m + m$) 
$\limp$ {\textquotesingle}\texttt{I\_}$m$ \texttt{:=} 
\vskip0pt\noindent\hskip10pt
  \texttt{if $m$ is $m_1$.+1 then fun $i$ => inZp ($i$./2) else fun $i$ => $i$}.
\vskip0pt\noindent\hskip0pt
\texttt{Definition} \textit{elift} $m$ \texttt{:} {\textquotesingle}\texttt{I\_}$m$ 
$\limp$ {\textquotesingle}\texttt{I\_}($m + m$) \texttt{:=} 
\vskip0pt\noindent\hskip10pt
  \texttt{if $m$ is $m_1$.+1 then fun $i$ => inZp ($i$.*2) else fun $i$ => $i$}.
\vskip0pt\noindent\hskip0pt
\texttt{Definition} \textit{olift} m \texttt{:} {\textquotesingle}\texttt{I\_}$m$
 $\limp$ {\textquotesingle}\texttt{I\_}($m$ + $m$) \texttt{:}= 
\vskip0pt\noindent\hskip10pt
  \texttt{if $m$ is $m_1$.+1 then fun $i$ => inZp ($i$.*2.+1) else fun $i$ => $i$}.
\end{framed}
\vskip0pt\noindent
Then, we can introduce the parity merge for connectors.
\vskip0pt\noindent
\noindent
\begin{framed}
\footnotesize
\noindent
\vskip0pt\noindent\hskip0pt
\texttt{Definition} \textit{clink\_eomerge} $m$ ($c_1$ \texttt{:} \textit{connector} $m$) 
($c_2$ \texttt{:} \textit{connector} $m$) \texttt{:=}
\vskip0pt\noindent\hskip10pt
  \texttt{[ffun $i$ : {\textquotesingle}{I\_}($m + m$) =>}
\vskip0pt\noindent\hskip20pt
    \texttt{if odd $i$ then} \textit{olift} (\textit{clink} $c_2$ 
    (\textit{idiv2} $i$))
\vskip0pt\noindent\hskip65pt
\texttt{else} \textit{elift} (\textit{clink} $c_1$
    (\textit{idiv2} $i$))\texttt{]}.
\vskip0pt\noindent\hskip0pt
\texttt{Lemma} \textit{clink\_eomerge\_proof} $m$ ($c_1$ \texttt{:} 
\textit{connector} $m$) ($c_2$ \texttt{:} \textit{connector} $m$) \texttt{:}
\vskip0pt\noindent\hskip10pt
  \texttt{[forall} $i$, (\textit{clink\_eomerge} $c_1$ $c_2$
   (\textit{clink\_eomerge} $c_1$ $c_2$ $i$)) \texttt{==} $i$\texttt{]}.
\vskip0pt\noindent\hskip0pt
\texttt{Definition} \textit{ceomerge} $m$ ($c_1$ \texttt{:} 
\textit{connector} $m$) ($c_2$ \texttt{:} \textit{connector} $m$) \texttt{:=} 
\vskip0pt\noindent\hskip10pt
  \textit{connector\_of} (\textit{clink\_eomerge\_proof} $c_1$ $c_2$).
\end{framed}
\vskip0pt\noindent
\noindent
Finally we can get the parity duplication
\begin{framed}
\footnotesize
\noindent
\vskip0pt\noindent\hskip0pt
\texttt{Definition} \textit{neomerge} $m$ ($n_1$ \texttt{:} 
\textit{network} $m$) ($n_2$ \texttt{:} \textit{network} $m$) \textit{:=} 
\vskip0pt\noindent\hskip10pt
    \texttt{[seq} \textit{ceomerge} $i$\texttt{.1} $i$\texttt{.2} 
    \texttt{|} $i$ \texttt{<-} \texttt{zip} $n_1$ $n_2$\texttt{]}.
\vskip0pt\noindent\hskip0pt
\texttt{Definition} \textit{neodup} $m$ ($n$ \texttt{:} \textit{network} $m$) : 
\textit{network} ($m + m$) \texttt{:=} \textit{neomerge} $n$ $n$.
\end{framed}
\vskip0pt\noindent

\section{Bitonic Sorter}

Here is the version of the bitonic sorter for 8 lines.
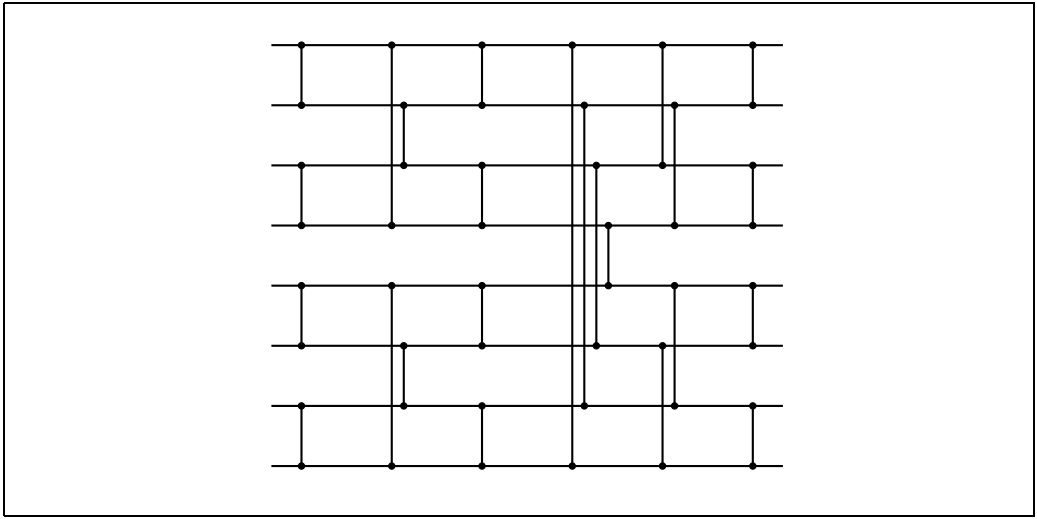
\begin{boxed}
\vspace{5pt}
\vskip5pt
\begin{center}
\begin{tikzpicture}[scale=0.80]
\foreach \a in {1,...,8}
  \draw[thick] (0,\a) -- ++(8.5,0);
\foreach \x in {    
{0.5,1},
{0.5,2},
{0.5,3},
{0.5,4},
{0.5,5},
{0.5,6},
{0.5,7},
{0.5,8},
{2,1},
{2,4},
{2,5},
{2,8},
{2.2,2},
{2.2,3},
{2.2,6},
{2.2,7},
{3.5,1},
{3.5,2},
{3.5,3},
{3.5,4},
{3.5,5},
{3.5,6},
{3.5,7},
{3.5,8},
{5,1},
{5.2,2},
{5.4,3},
{5.6,4},
{5.6,5},
{5.4,6},
{5.2,7},
{5,8},
{6.5,1},
{6.5,3},
{6.5,8},
{6.5,6},
{6.7,2},
{6.7,4},
{6.7,7},
{6.7,5},
{8,1},
{8,2},
{8,3},
{8,4},
{8,5},
{8,6},
{8,7},
{8,8}}
  \filldraw (\x) circle (1.5pt);
\draw[thick] (0.5,1) -- (0.5,2);
\draw[thick] (0.5,3) -- (0.5,4);
\draw[thick] (0.5,5) -- (0.5,6);
\draw[thick] (0.5,7) -- (0.5,8);
\draw[thick] (2,1) -- (2,4);
\draw[thick] (2,5) -- (2,8);
\draw[thick] (2.2,2) -- (2.2,3);
\draw[thick] (2.2,6) -- (2.2,7);
\draw[thick] (3.5,1) -- (3.5,2);
\draw[thick] (3.5,3) -- (3.5,4);
\draw[thick] (3.5,5) -- (3.5,6);
\draw[thick] (3.5,7) -- (3.5,8);
\draw[thick] (5,1) -- (5,8);
\draw[thick] (5.2,2) -- (5.2,7);
\draw[thick] (5.4,3) -- (5.4,6);
\draw[thick] (5.6,4) -- (5.6,5);
\draw[thick] (6.5,1) -- (6.5,3);
\draw[thick] (6.5,6) -- (6.5,8);
\draw[thick] (6.7,2) -- (6.7,4);
\draw[thick] (6.7,5) -- (6.7,7);
\draw[thick] (8,1) -- (8,2);
\draw[thick] (8,3) -- (8,4);
\draw[thick] (8,5) -- (8,6);
\draw[thick] (8,7) -- (8,8);
\end{tikzpicture}
\end{center}
\vskip5pt
\end{boxed}
\noindent
It is composed of 6 connectors (the drawing of some links have been 
slightly shifted to the right so they don't overlap). The key ingredient
of this network is the half-cleaner. It is a connector for $m + m$ lines, that
links the line $i$ to the line $i+m$ for $i < m$.
\begin{framed}
\footnotesize
\noindent
\vskip0pt\noindent\hskip0pt
\texttt{Definition} \textit{clink\_half\_cleaner} $m$ \texttt{:} 
\texttt{\{ffun} \texttt{I\_}($m$ + $m)$ $\limp$ \texttt{I\_}($m$ + $m$)
\texttt{\}} 
\texttt{:=}
\vskip0pt\noindent\hskip10pt
  \texttt{[ffun} $i$ \texttt{=>}
\vskip0pt\noindent\hskip20pt
    \texttt{match split} $i$ \texttt{with} 
\vskip0pt\noindent\hskip20pt
    \texttt{| inl} $x$ \texttt{=>} \texttt{rshift} \texttt{\_} $x$
\vskip0pt\noindent\hskip20pt
    \texttt{| inr} $x$ \texttt{=>} \texttt{lshift} \texttt{\_} $x$
\vskip0pt\noindent\hskip20pt
    \texttt{end]}.
\vskip0pt\noindent\hskip0pt
\texttt{Lemma} \textit{clink\_half\_cleaner\_proof} $m$ \texttt{:}
\vskip0pt\noindent\hskip10pt
  \texttt{[forall} $i$ : \texttt{I\_}($m + m$), \texttt{clink\_half\_cleaner}
  \texttt{\_} (\textit{clink\_half\_cleaner} \_ $i$) \texttt{==} $i$\texttt{]}.
\vskip0pt\noindent\hskip0pt
\texttt{Definition} \textit{half\_cleaner} $m$ \texttt{:=}
\textit{connector\_of} (\textit{clink\_half\_cleaner\_proof} $m$).
\end{framed}
\vskip0pt\noindent
This connector has an interesting behaviour when given as input a so-called
bitonic tuple. Technically, a sequence of elements is bitonic if there is 
one of its rotation that is increasing then decreasing.
\begin{framed}
\footnotesize
\noindent
\vskip0pt\noindent\hskip0pt
\texttt{Definition} \textit{bitonic} \texttt{:=} \texttt{[qualify} $s$
 \texttt{|} 
\vskip0pt\noindent\hskip10pt
 \texttt{[exists} $r$ \texttt{:} \texttt{I\_}(\texttt{size} $s$)\texttt{.+1}, 
\vskip0pt\noindent\hskip15pt
  \texttt{exists} $n$ \texttt{:} \texttt{I\_}(\texttt{size} $s$).\texttt{+1},
\vskip0pt\noindent\hskip15pt
  \texttt{let} $s_1$ \texttt{:=} \texttt{rot} $r$ $s$ \texttt{in} \texttt{sorted}
   \texttt{$\le$} (\texttt{take} $n$ $s_1$) \texttt{\&\&} \texttt{sorted} \texttt{$\ge$}
  (\texttt{drop} $n$ $s_1$)\texttt{]]}.
\end{framed}
\vskip0pt\noindent
Fortunately for sequences of booleans the characterisation is simpler : 
a sequence of booleans is bitonic if it has at most 2 flips.
\begin{framed}
\footnotesize
\noindent
\vskip0pt\noindent\hskip0pt
\texttt{Lemma} \textit{bitonic\_boolP} ($s$ : \texttt{seq bool}) \texttt{:}
\vskip0pt\noindent\hskip10pt
  \texttt{reflect} (\texttt{exists} $t$,
\vskip0pt\noindent\hskip60pt
            \texttt{let:} ($b$,$i$,$j$,$k$) \texttt{:}= $t$ \texttt{in}
             $s$ = \texttt{nseq} $i$ $b$ \texttt{++}
                  \texttt{nseq} $j$ (\texttt{\textctilde\textctilde}\,$b$) \texttt{++}
                   \texttt{nseq} $k$ $b$)
\vskip0pt\noindent\hskip50pt
          ($s$ \texttt{is} \textit{bitonic}).
\end{framed}
\vskip0pt\noindent
When applied to a bitonic sequence, the half-cleaner returns a tuple 
whose right half contains only \texttt{true} and the left half is bitonic 
or the left half contains only \texttt{false} and the right half is 
bitonic.
\begin{framed}
\footnotesize
\noindent
\vskip0pt\noindent\hskip0pt
\texttt{Lemma} \textit{bitonic\_half\_cleaner} $m$ ($t$ : ($m + m$)\texttt{.-tuple}
 \texttt{bool}) \texttt{:}
\vskip0pt\noindent\hskip10pt
  $t$ \texttt{is} \textit{bitonic} $\limp$ 
\vskip0pt\noindent\hskip10pt
  \texttt{let} $t_1$ \texttt{:=} \texttt{cfun} (\textit{half\_cleaner} $m$) $t$ 
  \texttt{in} 
\vskip0pt\noindent\hskip20pt
    ((\texttt{take} $m$ $t_1$ \texttt{==} \texttt{nseq} $n$ \texttt{false}) 
    \texttt{\&\&} (\texttt{drop} $m$ $t_1$ \texttt{is} 
    \textit{bitonic}))
\vskip0pt\noindent\hskip10pt
  \texttt{||}
\vskip0pt\noindent\hskip20pt
    ((\texttt{drop} $m$ $t_1$ \texttt{==} \texttt{nseq} $n$ \texttt{true}) 
    \texttt{\&\&} (\texttt{take} $m$ $t_1$ \texttt{is} 
    \textit{bitonic})).
\end{framed}
\vskip0pt\noindent
The proof proceeds by case analysis. As the tuple contains only 2 flips,
there are two easy cases when these two flips are both in a single half.
When it is in the left half, we have
$$
\arraycolsep=10.4pt\def\arraystretch{1.2}
\begin{array}{|r|r|r|r|r|r|r|r|r|r|r|}
\hline
\textrm{left half}  & b & b & b & \overline{b} & \overline{b} & 
 \overline{b} & \overline{b} & b & b & b \cr 
\hline
\textrm{right half}  & b & b & b & b & b & b & b & b & b & b  \cr
\hline
\textrm{min}  & b & b & b & \texttt{F} & \texttt{F} & \texttt{F} & \texttt{F} & b & b & b  \cr 
\hline
\textrm{max}  & b & b & b & \texttt{T} & \texttt{T} & \texttt{T} & \texttt{T} & b & b & b  \cr
\hline
\end{array}
$$
so the property holds. By symmetry this is the same if the two flips 
are on right half.
$$
\arraycolsep=10.4pt\def\arraystretch{1.2}
\begin{array}{|r|r|r|r|r|r|r|r|r|r|r|}
\hline
\textrm{left half}  & b & b & b & b & b & b & b & b & b & b  \cr
\hline
\textrm{right half}  & b & b & b & \overline{b} & \overline{b} & 
 \overline{b} & \overline{b} & b & b & b \cr 
\hline
\textrm{min}  & b & b & b & \texttt{F} & \texttt{F} & \texttt{F} & \texttt{F} & b & b & b  \cr 
\hline
\textrm{max}  & b & b & b & \texttt{T} & \texttt{T} & \texttt{T} & \texttt{T} & b & b & b  \cr
\hline
\end{array}
$$
In the remaining cases, each half has a flip. Suppose the flip in the 
left half occurs first, we have:
$$
\arraycolsep=10.4pt\def\arraystretch{1.2}
\begin{array}{|r|r|r|r|r|r|r|r|r|r|r|}
\hline
\textrm{left half}  & b & b & b & \overline{b} & \overline{b} & \overline{b} 
    & \overline{b} & \overline{b} & \overline{b} & \overline{b}  \cr
\hline
\textrm{right half}  & \overline{b} & \overline{b} & \overline{b} & \overline{b} & \overline{b} & 
 \overline{b} & \overline{b} & b & b & b \cr 
\hline
\textrm{min}  & \texttt{F} & \texttt{F} & \texttt{F} & \overline{b} & \overline{b} & \overline{b} & 
\overline{b} & \texttt{F} & \texttt{F} & \texttt{F}  \cr 
\hline
\textrm{max}  & \texttt{T} & \texttt{T} & \texttt{T} & \overline{b} & \overline{b} & \overline{b} 
& \overline{b} & \texttt{T} & \texttt{T} & \texttt{T}  \cr
\hline
\end{array}
$$
and the property holds again.
Finally the flip in the right half occurs first, we have
$$
\arraycolsep=10.4pt\def\arraystretch{1.2}
\begin{array}{|r|r|r|r|r|r|r|r|r|r|r|}
\hline
\textrm{left half}  & b & b & b & b & b & b 
    & b & \overline{b} & \overline{b} & \overline{b}  \cr
\hline
\textrm{right half}  & \overline{b} & \overline{b} & \overline{b} & b & b & 
 b & b & b & b & b \cr 
\hline
\textrm{min}  & \texttt{F} & \texttt{F} & \texttt{F} & b & b & b & 
b & \texttt{\texttt{F}} & \texttt{F} & \texttt{F}  \cr 
\hline
\textrm{max}  & \texttt{T} & \texttt{T} & \texttt{T} & b & b & b 
& b & \texttt{T} & \texttt{T} & \texttt{T}  \cr
\hline
\end{array}
$$
This ends the proof.

The next observation is that if we recursively apply on the 
resulting halves the half-cleaner, we end up getting a sorted
list : we progressively add \texttt{false} on the left part or 
\texttt{true} on 
the right one. Being able to perform this recursion on halves
implies that the initial number of lines must be a power of 2.
In our case, in order to insert a half-cleaner
we need to have a type of the form \textit{connector} ($m + m$).
This means that it is mandatory for the typechecker to succeed
that $2^{m+1}$ converts to $2^m + 2^m$. This is not the case
with the exponential function of the library. So we define our own 
version that we write $`2^n$ in the following.
\begin{framed}
\footnotesize
\noindent
\vskip0pt\noindent\hskip0pt
\texttt{Fixpoint} $`2^m$ \texttt{:=} 
\texttt{if} $m$ \texttt{is} $m_1$\texttt{.+1} 
\texttt{then} $`2^{m_1} + `2^{m_1}$ \texttt{else} 1.
\end{framed}
\vskip0pt\noindent
We can then define the recursive function.
\begin{framed}
\footnotesize
\noindent
\vskip0pt\noindent\hskip0pt
\texttt{Fixpoint} \textit{half\_cleaner\_rec} $m$ \texttt{:} 
\textit{network} $`2^ m$ \texttt{:=}
\vskip0pt\noindent\hskip10pt
  \texttt{if} $m$ \texttt{is} $m_1$\texttt{.+1} 
  \texttt{then} \textit{half\_cleaner} $`2^{m_1}$ \texttt{::}
  \textit{ndup} (\textit{half\_cleaner\_rec} $m_1$)
\vskip0pt\noindent\hskip10pt
  \texttt{else} \texttt{[::]}.
\end{framed}
\vskip0pt\noindent
We can then easily prove its expected behaviour.
\begin{framed}
\footnotesize
\vskip0pt\noindent
\texttt{Lemma} \textit{sorted\_half\_cleaner\_rec} $m$ ($t$ : $`2^m$
\texttt{.-tuple} \texttt{bool}) \texttt{:}
\vskip0pt\noindent\hskip10pt
  $t$ \texttt{is} \textit{bitonic} $\limp$
  \texttt{sorted} \texttt{$\le$} (\textit{nfun} (\textit{half\_cleaner\_rec} $m$)
   $t$).
\end{framed}
\vskip0pt\noindent
and show that it is logarithmic and creates a network of $m$ connectors.
\begin{framed}
\footnotesize
\noindent
\vskip0pt\noindent\hskip0pt
\texttt{Lemma} \textit{size\_half\_cleaner\_rec} $m$ : \texttt{size}
 (\textit{half\_cleaner\_rec} $m$) = $m$.
\vskip0pt\noindent
\end{framed}
\vskip0pt\noindent
The recursive half-cleaner requires to have a bitonic entry.
If we try to build a recursive algorithm, calling it 
first on the top-half lines and then on the bottom-half lines, we get 
two sorting outputs.
Gluing them directly does not give a bitonic entry.
There are possibly too many flips.
Each half that is sorted contains potentially a flip and there is
the potential flip at their intersection. Instead, the trick 
is to glue them together but reversing the second one.
This leads to a bitonic entry. So,
a reverse version of the half-cleaner is created that performs 
this reversal. Graphically, it looks like this.
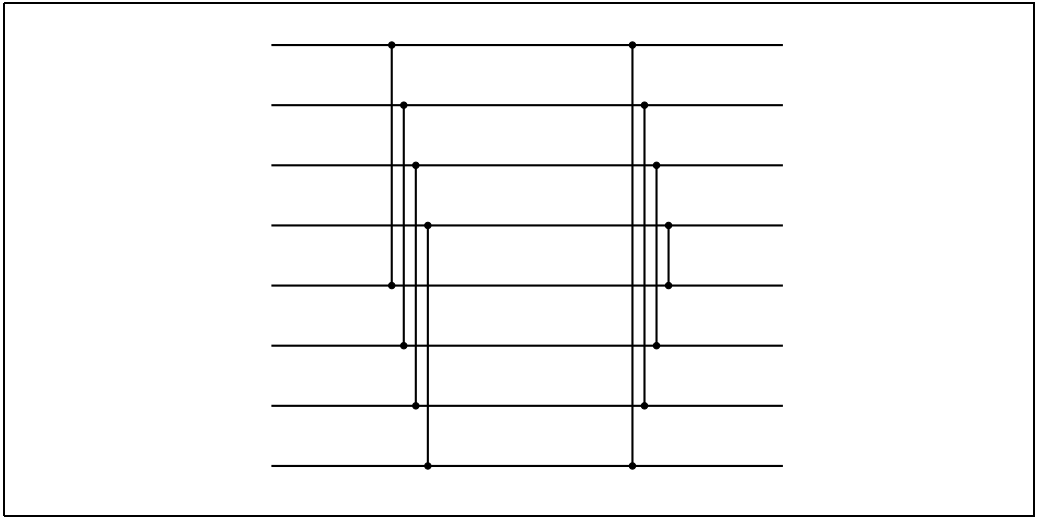
\begin{boxed}
\vspace{5pt}
\vskip5pt
\begin{center}
\begin{tikzpicture}[scale=0.80]
\foreach \a in {1,...,8}
  \draw[thick] (0,\a) -- ++(8.5,0);
\foreach \x in {    
{2,4},
{2,8},
{2.2,3},
{2.2,7},
{2.4,2},
{2.4,6},
{2.6,1},
{2.6,5},
{6,1},
{6.2,2},
{6.4,3},
{6.6,4},
{6.6,5},
{6.4,6},
{6.2,7},
{6,8}}
  \filldraw (\x) circle (1.5pt);
\draw[thick] (2,4) -- (2,8);
\draw[thick] (2.2,3) -- (2.2,7);
\draw[thick] (2.4,2) -- (2.4,6);
\draw[thick] (2.6,1) -- (2.6,5);
\draw[thick] (6,1) -- (6,8);
\draw[thick] (6.2,2) -- (6.2,7);
\draw[thick] (6.4,3) -- (6.4,6);
\draw[thick] (6.6,4) -- (6.6,5);
\end{tikzpicture}
\end{center}
\vskip5pt
\end{boxed}
\noindent
On the left-hand side there is the standard half-cleaner. On the 
right-hand side there is the reverse version where the link to 
the bottom lines have been reverse. For example,
the line 1 is linked to the line 5 on the left part.
It is now linked to the line 8 on the right part.
There is a \texttt{rev\_ord} function for ordinals. We use it 
to implement the reverse half-cleaner, so the line $i$ is connected
to line $m - i$:
\begin{framed}
\footnotesize
\vskip0pt\noindent
\texttt{Definition} \textit{clink\_rhalf\_cleaner} $m$ : 
\texttt{\{ffun} 
\texttt{I\_}$m$ $\limp$ \texttt{I\_}$m$\texttt{\}} \texttt{:=}
\texttt{[ffun} $i$ \texttt{=>} \texttt{rev\_ord} $i$\texttt{]}.
\vskip0pt\noindent
\texttt{Lemma} \textit{clink\_rhalf\_cleaner\_proof} $m$ \texttt{:} 
\vskip0pt\noindent\hskip10pt
  \texttt{[forall} $i$ \texttt{:} \texttt{I\_}($m + m$), 
  \textit{clink\_rhalf\_cleaner} \_ (\textit{clink\_rhalf\_cleaner} \_ $i$) 
  \texttt{==} $i$\texttt{]}.
\vskip0pt\noindent\hskip0pt
\texttt{Definition} \textit{rhalf\_cleaner} $m$\texttt{:=} 
\textit{connector\_of} (\textit{clink\_rhalf\_cleaner\_proof} $m$).
\vskip0pt\noindent\hskip0pt
\end{framed}
\vskip0pt\noindent
Now, we can use the reverse half-cleaner before calling 
the recursive half-cleaner.
\begin{framed}
\footnotesize
\vskip0pt\noindent
\texttt{Definition} \textit{rhalf\_cleaner\_rec} $n$ \texttt{:}
 \textit{network} `$2^n$ \texttt{:=}
\vskip0pt\noindent\hskip10pt
  \texttt{if} $n$ \texttt{is} $n_1$\texttt{.+1} \texttt{then}
    \textit{rhalf\_cleaner} `$2^{n_1}$ \texttt{::} \textit{ndup}
     (\textit{half\_cleaner\_rec} $n_1$)
\vskip0pt\noindent\hskip10pt
  \texttt{else} \texttt{[::]}.
\end{framed}
\vskip0pt\noindent
The call to the reverse half-cleaner produces on the top-half lines either only 
\texttt{true} values so there is no problem or a reverse 
of a bitonic but it is also ok, the reverse of a bitonic is a 
bitonic. The same holds for the bottom-half lines.
So, we get the expected theorem. 
\begin{framed}
\footnotesize
\vskip0pt\noindent
\texttt{Lemma} \textit{sorted\_rhalf\_cleaner\_rec} $m$ ($t$ \texttt{:}
 `$2^{m.+1}$\texttt{.-tuple} \texttt{bool}) \texttt{:}
\vskip0pt\noindent\hskip10pt
  \texttt{sorted} \texttt{$\le$} (\texttt{take} `$2^{m}$ $t$) 
  $\limp$ \texttt{sorted} \texttt{$\le$} (\texttt{drop} `$2^ m$ $t$) 
  $->$
\vskip0pt\noindent\hskip10pt
  \texttt{sorted} \texttt{$\le$} (\textit{nfun} (\textit{rhalf\_cleaner\_rec} $m$\texttt{.+1})
   $t$).
\end{framed}
\vskip0pt\noindent
Now, we can build the recursion
\begin{framed}
\footnotesize
\vskip0pt\noindent
\texttt{Fixpoint} \textit{bsort} $m$ \texttt{:}
\textit{network} `$2^ m$ \texttt{:=}
\vskip0pt\noindent\hskip10pt
  \texttt{if} \texttt{m} \texttt{is} $m_1$\texttt{.+1}
  \texttt{then} \textit{ndup} (\textit{bsort} $m_1$) \texttt{++}
   \textit{rhalf\_cleaner\_rec} $m_1$\texttt{.+1} 
\vskip0pt\noindent\hskip10pt
  \texttt{else} \texttt{[::]}.
\end{framed}
\vskip0pt\noindent
and get the final results.
\begin{framed}
\footnotesize
\vskip0pt\noindent
\texttt{Lemma} \textit{sorting\_bsort} $m$ : \textit{bsort}
 $m$ \texttt{is} \textit{sorting}.
\vskip0pt\noindent\hskip0pt
\texttt{Lemma} \textit{size\_bsort} $m$ \texttt{:}
 \texttt{size} (\textit{bsort} $m$) = 
 ($m$ $*$ $m$\texttt{.+1})\texttt{./2}.
\end{framed}
\vskip0pt\noindent
Here is the complete code of the algorithm.
\begin{framed}
\footnotesize
\vskip0pt\noindent\hskip0pt
\texttt{Fixpoint} \textit{half\_cleaner\_rec} $m$ \texttt{:} 
\textit{network} $`2^ m$ \texttt{:=}
\vskip0pt\noindent\hskip10pt
  \texttt{if} $m$ \texttt{is} $m_1$\texttt{.+1} 
  \texttt{then} \textit{half\_cleaner} $`2^{m_1}$ \texttt{::}
  \textit{ndup} (\textit{half\_cleaner\_rec} $m_1$)
\vskip0pt\noindent\hskip10pt
  \texttt{else} \texttt{[::]}.
\vskip0pt\noindent\hskip0pt
\texttt{Definition} \textit{rhalf\_cleaner\_rec} $n$ \texttt{:}
 \textit{network} `$2^n$ \texttt{:=}
\vskip0pt\noindent\hskip10pt
  \texttt{if} $n$ \texttt{is} $n_1$\texttt{.+1} \texttt{then}
    \textit{rhalf\_cleaner} `$2^{n_1}$ \texttt{::} \textit{ndup}
     (\textit{half\_cleaner\_rec} $n_1$)
\vskip0pt\noindent\hskip10pt
  \texttt{else} \texttt{[::]}.
\vskip0pt\noindent\hskip0pt
\texttt{Fixpoint} \textit{bsort} $m$ \texttt{:}
\textit{network} `$2^ m$ \texttt{:=}
\vskip0pt\noindent\hskip10pt
  \texttt{if} \texttt{m} \texttt{is} $m_1$\texttt{.+1}
  \texttt{then} \textit{ndup} (\textit{bsort} $m_1$) \texttt{++}
   \textit{rhalf\_cleaner\_rec} $m_1$\texttt{.+1} 
\vskip0pt\noindent\hskip10pt
  \texttt{else} \texttt{[::]}.
\end{framed}
\vskip0pt\noindent

\section{Knuth's Exchange Odd Even Sorter}

Here is the drawing of the odd-even sorter.
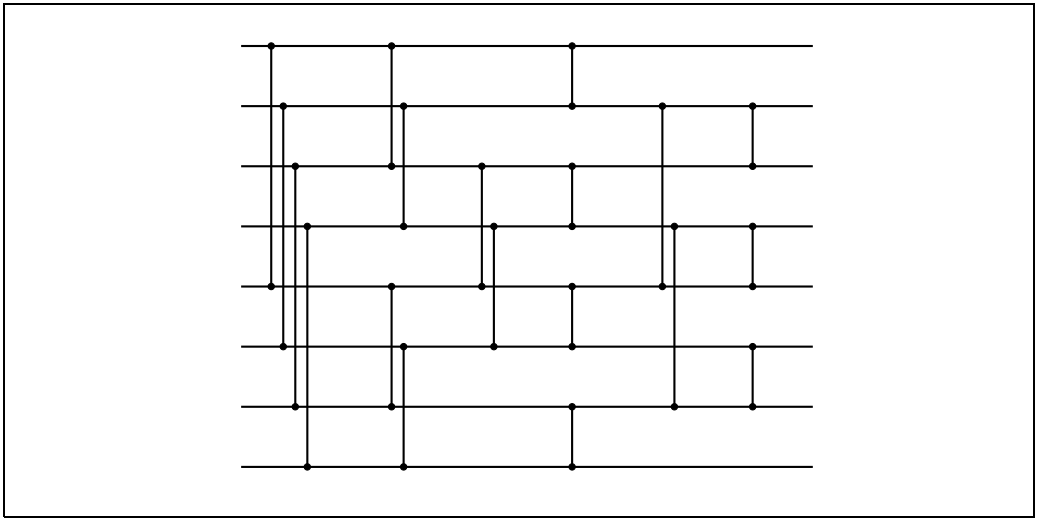
\begin{boxed}
\vspace{5pt}
\vskip5pt
\begin{center}
\begin{tikzpicture}[scale=0.80]
\foreach \a in {1,...,8}
  \draw[thick] (0,\a) -- ++(9.5,0);
\foreach \x in {    
{0.5,8},
{0.5,4},
{0.7,7},
{0.7,3},
{0.9,6},
{0.9,2},
{1.1,5},
{1.1,1},
{2.5,8},
{2.5,6},
{2.7,7},
{2.7,5},
{2.5,4},
{2.5,2},
{2.7,3},
{2.7,1},
{4,6},
{4,4},
{4.2,5},
{4.2,3},
{5.5,8},
{5.5,7},
{5.5,6},
{5.5,5},
{5.5,4},
{5.5,3},
{5.5,2},
{5.5,1},
{7,7},
{7,4},
{7.2,5},
{7.2,2},
{8.5,7},
{8.5,6},
{8.5,5},
{8.5,4},
{8.5,3},
{8.5,2}}
  \filldraw (\x) circle (1.5pt);
\draw[thick] (0.5,8) -- (0.5,4);
\draw[thick] (0.7,7) -- (0.7,3);
\draw[thick] (0.9,6) -- (0.9,2);
\draw[thick] (1.1,5) -- (1.1,1);
\draw[thick] (2.5,8) -- (2.5,6);
\draw[thick] (2.7,7) -- (2.7,5);
\draw[thick] (2.5,4) -- (2.5,2);
\draw[thick] (2.7,3) -- (2.7,1);
\draw[thick] (4,6) -- (4,4);
\draw[thick] (4.2,5) -- (4.2,3);
\draw[thick] (5.5,8) -- (5.5,7);
\draw[thick] (5.5,6) -- (5.5,5);
\draw[thick] (5.5,4) -- (5.5,3);
\draw[thick] (5.5,2) -- (5.5,1);
\draw[thick] (7,7) -- (7,4);
\draw[thick] (7.2,5) -- (7.2,2);
\draw[thick] (8.5,7) -- (8.5,6);
\draw[thick] (8.5,5) -- (8.5,4);
\draw[thick] (8.5,3) -- (8.5,2);
\end{tikzpicture}
\end{center}
\vskip5pt
\end{boxed}
\noindent
This is still a recursive algorithm but this time it is not based 
on a top-half, bottom-half partition but an even and odd partition.
We add them as basic operations on sequences.
\begin{framed}
\footnotesize
\vskip0pt\noindent
\texttt{Fixpoint} \textit{etake} $s$ \texttt{:=}
\vskip0pt\noindent\hskip10pt
  \texttt{if} $s$ \texttt{is} $a$ :: $s_1$ \texttt{then} $a$ :: (\texttt{if}
  $s_1$ \texttt{is} \texttt{\_} \texttt{::} $s_2$ \texttt{then} \textit{etake} $s_2$ \texttt{else}
   \texttt{[::]})
\vskip0pt\noindent\hskip10pt
  \texttt{else} \texttt{[::]}.
\vskip0pt\noindent\hskip0pt
\texttt{Definition} \textit{otake} $s$ \texttt{:=}
  \texttt{if} $s$ \texttt{is} \texttt{\_} \texttt{::} $s_1$ \texttt{then} \textit{etake} $s_1$ 
  \texttt{else} \texttt{[::]}.
\end{framed}
\vskip0pt\noindent
There are two components of this sorter. The first one 
is the one that connects even line to one of their odd neighbour.
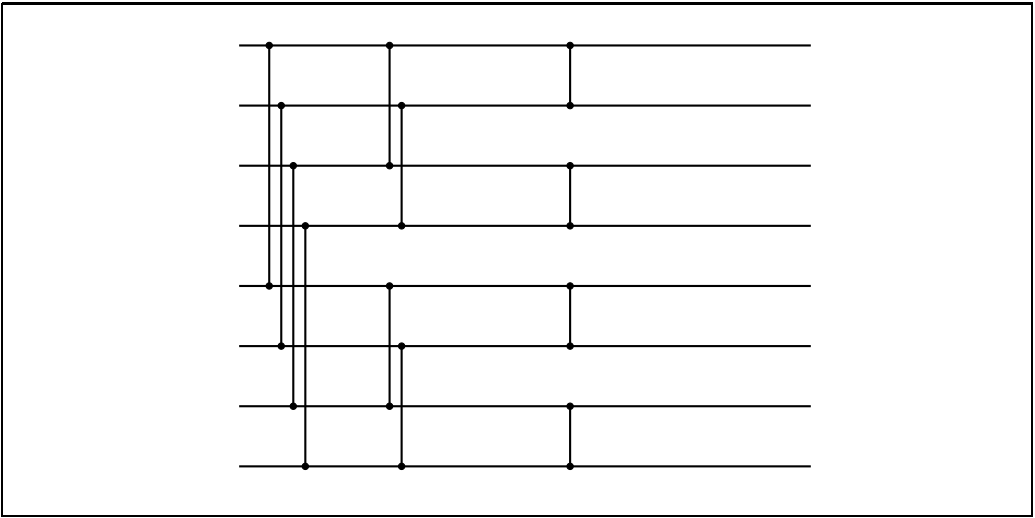
\begin{boxed}
\vspace{5pt}
\vskip5pt
\begin{center}
\begin{tikzpicture}[scale=0.80]
\foreach \a in {1,...,8}
  \draw[thick] (0,\a) -- ++(9.5,0);
\foreach \x in {    
{0.5,8},
{0.5,4},
{0.7,7},
{0.7,3},
{0.9,6},
{0.9,2},
{1.1,5},
{1.1,1},
{2.5,8},
{2.5,6},
{2.7,7},
{2.7,5},
{2.5,4},
{2.5,2},
{2.7,3},
{2.7,1},
{5.5,8},
{5.5,7},
{5.5,6},
{5.5,5},
{5.5,4},
{5.5,3},
{5.5,2},
{5.5,1}}
  \filldraw (\x) circle (1.5pt);
\draw[thick] (0.5,8) -- (0.5,4);
\draw[thick] (0.7,7) -- (0.7,3);
\draw[thick] (0.9,6) -- (0.9,2);
\draw[thick] (1.1,5) -- (1.1,1);
\draw[thick] (2.5,8) -- (2.5,6);
\draw[thick] (2.7,7) -- (2.7,5);
\draw[thick] (2.5,4) -- (2.5,2);
\draw[thick] (2.7,3) -- (2.7,1);
\draw[thick] (5.5,8) -- (5.5,7);
\draw[thick] (5.5,6) -- (5.5,5);
\draw[thick] (5.5,4) -- (5.5,3);
\draw[thick] (5.5,2) -- (5.5,1);
\end{tikzpicture}
\end{center}
\vskip5pt
\end{boxed}
\noindent
We first have 4 copies with jump 4 then 2 copies with jump 2 finally 1 copy
with jump 1.
The copy with jump 1 on the right shows the structure: even lines are linked
to their down neighbour. 
In order to encode it, we need to introduce the notion of neighbour 
for ordinals.
\begin{framed}
\footnotesize
\vskip0pt\noindent\hskip0pt
\texttt{Definition} \textit{inext} $m$ \texttt{:} 
\texttt{I\_}$m$ \texttt{->} \texttt{I\_}$m$ := 
\vskip0pt\noindent\hskip10pt
  \texttt{if} $m$ \texttt{is} $m_1$\texttt{.+1} \texttt{then}
   \texttt{fun} $i$ \texttt{=>} \texttt{inZp}
    (\texttt{if} $i$ \texttt{==} $m_1$ \texttt{then} $i$ 
    \texttt{else} $i$\texttt{.+1})
\vskip0pt\noindent\hskip10pt
  \texttt{else} \texttt{fun} $i$ \texttt{=>} $i$.
\vskip0pt\noindent\hskip0pt
\texttt{Definition} \textit{ipred} $m$ \texttt{:}
 \texttt{I\_}$m$ \texttt{->} \texttt{I\_}$m$ $:=$ 
\vskip0pt\noindent\hskip10pt
\texttt{if} $m$ \texttt{is} $m_1$\texttt{.+1}
 \texttt{then} \texttt{fun} $i$ \texttt{=>}
  \texttt{inZp} ($i$\texttt{.-1}) \texttt{else} \texttt{fun} $i$ 
  \texttt{=>} $i$.
\end{framed}
\vskip0pt\noindent
We can define the connector.
\begin{framed}
\footnotesize
\vskip0pt\noindent\hskip0pt
\texttt{Definition} \textit{clink\_eswap} $m$\texttt{:}
 \texttt{\{ffun} \texttt{I\_}$m$ \texttt{->} \texttt{I\_}$m$\texttt{\}} \texttt{:=}
 \vskip0pt\noindent\hskip10pt
\texttt{[ffun} $i$ \texttt{:} \texttt{I\_} \_ \texttt{=>}
 \texttt{if} \texttt{odd} $i$ \texttt{then} \textit{ipred} $i$
\texttt{else} \textit{inext} $i$\texttt{]}.
\vskip0pt\noindent\hskip0pt
\texttt{Lemma} \textit{clink\_eswap\_proof} $m$ \texttt{:} 
\vskip0pt\noindent\hskip10pt
  \texttt{[forall} $i$ \texttt{:} \texttt{I\_}$m$, 
  \textit{clink\_eswap} \_ (\textit{clink\_eswap} \_ $i$) \texttt{==} $i$
  \texttt{]}.
\vskip0pt\noindent\hskip0pt
\texttt{Definition} \textit{ceswap} $m$ \texttt{:=}
 \textit{connector\_of} (\textit{clink\_eswap\_proof} $m$).
\end{framed}
\vskip0pt\noindent
If we look at the effect of applying this connector to a tuple  of
booleans,
if the even lines and the odd lines are sorted, this property 
is preserved plus the even part contains more \texttt{false}
than the odd part:
\begin{framed}
\footnotesize
\vskip0pt\noindent\hskip0pt
\texttt{Definition} \textit{noF} ($s$ : \texttt{seq} \texttt{bool}) 
\texttt{:=} \texttt{count} (\texttt{fun} $b$ \texttt{=>} 
\texttt{\textctilde\textctilde}$b$) $s$.
\vskip0pt\noindent\hskip0pt
\texttt{Lemma} \textit{sorted\_eswap} $m$ ($t$ : ($m + m$)\texttt{.-tuple}
 \texttt{bool}) \texttt{:}
\vskip0pt\noindent\hskip10pt
  \texttt{sorted}$_\le$ (\textit{etake} $t$) \texttt{->}
  \texttt{sorted}$_\le$ (\textit{otake} $t$) \texttt{->} 
\vskip0pt\noindent\hskip10pt
  \texttt{let} $t_1$\texttt{:=} \texttt{cfun} \textit{ceswap} $t$ \texttt{in}
\vskip0pt\noindent\hskip10pt
  \texttt{[$\land$} \texttt{sorted}$_\le$ (\textit{etake} $t_1$), 
\vskip0pt\noindent\hskip25pt
      \texttt{sorted}$_\le$ (\textit{otake} $t_1$) \texttt{\&} 
\vskip0pt\noindent\hskip25pt
      \textit{noF} (\textit{otake} $t_1$) $\le$ 
      \textit{noF} (\textit{etake} $t_1$)\texttt{]}.
\end{framed}
\vskip0pt\noindent
The second connector is the one that connects the odd lines 
with a $k$ jump ($k$ is odd) to the even lines.
\vskip0pt\noindent
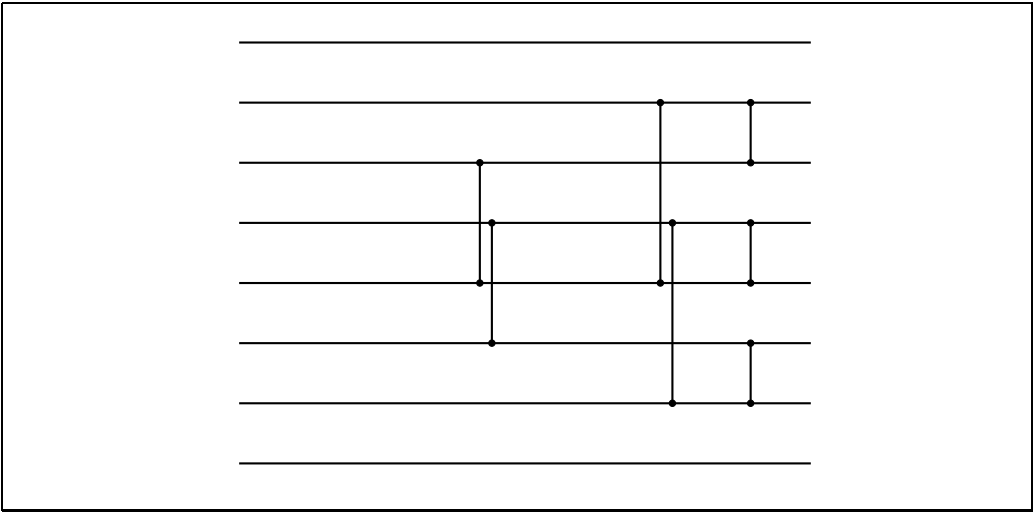
\begin{boxed}
\vspace{5pt}
\vskip5pt
\begin{center}
\begin{tikzpicture}[scale=0.80]
\foreach \a in {1,...,8}
  \draw[thick] (0,\a) -- ++(9.5,0);
\foreach \x in {    
{4,6},
{4,4},
{4.2,5},
{4.2,3},
{7,7},
{7,4},
{7.2,5},
{7.2,2},
{8.5,7},
{8.5,6},
{8.5,5},
{8.5,4},
{8.5,3},
{8.5,2}}
  \filldraw (\x) circle (1.5pt);
\draw[thick] (4,6) -- (4,4);
\draw[thick] (4.2,5) -- (4.2,3);
\draw[thick] (7,7) -- (7,4);
\draw[thick] (7.2,5) -- (7.2,2);
\draw[thick] (8.5,7) -- (8.5,6);
\draw[thick] (8.5,5) -- (8.5,4);
\draw[thick] (8.5,3) -- (8.5,2);
\end{tikzpicture}
\end{center}
\vskip5pt
\end{boxed}
\noindent
There are 2 copies with jump 1, then one copy with jump 3 and 
one copy with jump 1.
Again, we define first the operation on ordinals.
\begin{framed}
\footnotesize
\vskip0pt\noindent\hskip0pt
\texttt{Definition} \textit{iadd} $m$ $k$ \texttt{:} \texttt{I\_}$m$
 \texttt{->} \texttt{I\_}$m$ \texttt{:=}
\vskip0pt\noindent\hskip10pt
  \texttt{if} $m$ \texttt{is} $m_1$\texttt{.+1} \texttt{then}
   \texttt{fun} $i$ \texttt{=>} \texttt{inZp} (\texttt{if} $k + i$ $\le$ $m_1$
    \texttt{then} $k + i$ \texttt{else} $i$)
\vskip0pt\noindent\hskip10pt
  \texttt{else} \texttt{fun} $i$ \texttt{=>} $i$.
\vskip0pt\noindent\hskip0pt
\texttt{Definition} \textit{isub} $m$ $k$ \texttt{:}
 \texttt{I\_}$m$ \texttt{->} \texttt{I\_}$m$ \texttt{:=} 
\vskip0pt\noindent\hskip10pt
  \texttt{if} $m$ \texttt{is} $m_1$\texttt{.+1}
   \texttt{then} \texttt{fun} $i$ \texttt{=>}
    \texttt{inZp} (\texttt{if} $k \le i$ \texttt{then} $i - k$
    \texttt{else} $i$)
\vskip0pt\noindent\hskip10pt
  \texttt{else} \texttt{fun} $i$ \texttt{=>} $i$.
\end{framed}
\vskip0pt\noindent
We then create the connector. 
\begin{framed}
\footnotesize
\vskip0pt\noindent\hskip0pt
\texttt{Definition} \textit{clink\_odd\_jump} $m$ $k$ \texttt{:}
 \texttt{\{ffun} \texttt{I\_}$m$ \texttt{->} \texttt{I\_}$m$\texttt{\}} :=
\vskip0pt\noindent\hskip10pt
  \texttt{if} \texttt{odd} $k$ \texttt{then} 
    \texttt{[ffun} $i$ \texttt{=>}
     \texttt{if} \texttt{odd} $i$ \texttt{then}
      \textit{iadd} $k$ $i$ \texttt{else} \textit{isub} $k$ $i$
      \texttt{]}
\vskip0pt\noindent\hskip10pt
  \texttt{else} \texttt{[ffun} $i$ \texttt{=>} $i$\texttt{]}.
\vskip0pt\noindent\hskip0pt
\texttt{Lemma} \textit{clink\_odd\_jump\_proof} $m$ $k$ \texttt{:} 
\vskip0pt\noindent\hskip10pt
  \texttt{[forall} $i$ \texttt{:} \texttt{I\_}$m$, 
  \textit{clink\_odd\_jump} \_ $k$ (\textit{clink\_odd\_jump} \_ $k$ $i$) 
  \texttt{==} $i$\texttt{]}. 
\vskip0pt\noindent\hskip0pt
\texttt{Definition} \textit{codd\_jump {m}} $k$ \texttt{:=}
 \textit{connector\_of} (\textit{clink\_odd\_jump\_proof} $m$ $k$).
\end{framed}
\vskip0pt\noindent
This time, the \texttt{false} values are moving from the even lines
to the odd lines and we can quantify exactly how much. 
\begin{framed}
\footnotesize
\vskip0pt\noindent\hskip0pt
\texttt{Lemma} \textit{sorted\_odd\_jump} $m$ ($t$ : ($m + m$)\texttt{.-tuple}
 bool) $i$ $k$ \texttt{:}
\vskip0pt\noindent\hskip10pt
  \texttt{odd} $k$ \texttt{->} $i$ $<=$ (\texttt{uphalf} $k$).\texttt{*2} 
  \texttt{->}
\vskip0pt\noindent\hskip10pt
  \texttt{sorted}$_\le$ (\textit{etake} $t$) \texttt{->}
  \texttt{sorted}$_\le$ (\textit{otake} $t$) \texttt{->} 
\vskip0pt\noindent\hskip10pt
  \textit{noF} (\textit{etake t}) = \textit{noF} (\textit{otake}
   $t$) $+$ $i$ \texttt{->}
\vskip0pt\noindent\hskip10pt
  \texttt{let} $j$ \texttt{:=} $i$ $-$ \texttt{uphalf} $k$ \texttt{in}  
\vskip0pt\noindent\hskip10pt
  \texttt{let} $t_1$ \texttt{:=} \textit{cfun} (\textit{codd\_jump} $k$) 
  $t$ \texttt{in}
\vskip0pt\noindent\hskip20pt
  \texttt{[$\land$} \texttt{sorted}$_\le$ (\textit{etake} $t_1$), 
\vskip0pt\noindent\hskip35pt
      \texttt{sorted}$_\le$ (\textit{otake} $t_1$) \texttt{\&} 
\vskip0pt\noindent\hskip35pt
      \textit{noF} (\textit{etake} $t_1$) = 
      \textit{noF} (\textit{otake} $t_1$) $+$ ($i$ $-$ $j$\texttt{.*2})\texttt{]}.
\end{framed}
\noindent
Note that here we make use of the fact that $m - n = 0$ if $n \ge m$.

Now, the idea of the algorithm is to reduce the difference 
between the number of \texttt{false} between the odd and the 
even part so that the list becomes sorted.
\vskip0pt\noindent
\begin{framed}
\footnotesize
\vskip0pt\noindent\hskip0pt
\texttt{Lemma} \textit{sorted\_etake\_otake} $m$ ($t$ \texttt{:}
 ($m$ + $m$)\texttt{.-tuple} \texttt{bool}) \texttt{:}
\vskip0pt\noindent\hskip10pt
  \texttt{sorted}$_\le$ (\textit{etake} $t$) \texttt{->}
  \texttt{sorted}$_\le$ (\textit{otake} $t$) \texttt{->}
\vskip0pt\noindent\hskip10pt
  \textit{noF} (\textit{otake} $t$) $\le$
   \textit{noF} (\textit{etake} $t$) $\le$
    (\textit{noF} (\textit{otake} $t$))\texttt{.+1} \texttt{->}
\vskip0pt\noindent\hskip10pt
  \texttt{sorted}$_\le$ $t$.
\end{framed}
\vskip0pt\noindent
This is done by recursively halfing the jump and we get 
the expected result.
\vskip0pt\noindent
\begin{framed}
\footnotesize
\vskip0pt\noindent\hskip0pt
\texttt{Fixpoint} \textit{knuth\_jump\_rec} $m$ $k$ $r$ \texttt{:}
 \textit{network} $m$ \texttt{:=}
\vskip0pt\noindent\hskip10pt
  \texttt{if} $k$ \texttt{is} $k_1$\texttt{.+1} \texttt{then}
   \textit{codd\_jump} $r$ \texttt{::} \textit{knuth\_jump\_rec}
    $m$ $k_1$ (\texttt{uphalf} $r$)\texttt{.-1}
\vskip0pt\noindent\hskip10pt
  \texttt{else} \texttt{[::]}.
\vskip0pt\noindent\hskip0pt
\texttt{Lemma} \textit{sorted\_knuth\_jump\_rec} $m$ ($t$ : 
  ($m + m$)\texttt{.-tuple} \texttt{bool}) $k$ \texttt{:}
\vskip0pt\noindent\hskip10pt
  \texttt{sorted}$_\le$ (\textit{etake} $t$) \texttt{->}
  \texttt{sorted}$_\le$ (\textit{otake} $t$) \texttt{->}
\vskip0pt\noindent\hskip10pt
  \textit{noF} (\textit{otake} $t$) $\le$ \textit{noF} (\textit{etake} $t$) 
  $\le$ \textit{noF} (\textit{otake}) $+$ $`2^ k$ \texttt{->}
\vskip0pt\noindent\hskip10pt
  \texttt{sorted}$_\le$ (\textit{nfun} (\textit{knuth\_jump\_rec}
   ($m + m$) $k$ ($`2^ k$)\texttt{.-1)} $t$).
\end{framed}
\vskip0pt\noindent
We can now put together the recursion, the even swap and the 
recursive jump to get the sorter.
\vskip0pt\noindent
\begin{framed}
\footnotesize
\vskip0pt\noindent\hskip0pt
\texttt{Fixpoint} \textit{knuth\_exchange} $m$ \texttt{:}
 \textit{network} $`2^ m$ \texttt{:=}
\vskip0pt\noindent\hskip10pt
  \texttt{if} $m$ \texttt{is} $m_1$\texttt{.+1} \texttt{then}
\vskip0pt\noindent\hskip20pt
    \textit{neodup} (\textit{knuth\_exchange} $m_1$) \texttt{++}
     \textit{ceswap} \texttt{::} \textit{knuth\_jump\_rec} 
    $`2^ m$  $m_1$ (($`2^{m_1}$)\texttt{.-1})
\vskip0pt\noindent\hskip10pt
  \texttt{else} \texttt{[::]}.
\vskip0pt\noindent\hskip0pt
\texttt{Lemma} \textit{sorting\_knuth\_exchange} $m$ \texttt{:}
 \textit{knuth\_exchange} $m$ \texttt{is} \texttt{sorting}.
\vskip0pt\noindent\hskip0pt
\texttt{Lemma} \textit{size\_knuth\_exchange} $m$ \texttt{:}
 \texttt{size} (\textit{knuth\_exchange} $m$) = ($m * m$\texttt{.+1})\texttt{./2}.
\end{framed}
\vskip0pt\noindent
Here is the complete code of the algorithm.
\vskip0pt\noindent
\begin{framed}
\footnotesize
\vskip0pt\noindent\hskip0pt
\texttt{Fixpoint} \textit{knuth\_jump\_rec} $m$ $k$ $r$ \texttt{:}
 \textit{network} $m$ \texttt{:=}
\vskip0pt\noindent\hskip10pt
  \texttt{if} $k$ \texttt{is} $k_1$\texttt{.+1} \texttt{then}
   \textit{codd\_jump} $r$ \texttt{::} \textit{knuth\_jump\_rec}
    $m$ $k_1$ (\texttt{uphalf} $r$)\texttt{.-1}
\vskip0pt\noindent\hskip10pt
  \texttt{else} \texttt{[::]}.
\vskip0pt\noindent\hskip0pt
\texttt{Fixpoint} \textit{knuth\_exchange} $m$ \texttt{:}
 \textit{network} $`2^ m$ \texttt{:=}
\vskip0pt\noindent\hskip10pt
  \texttt{if} $m$ \texttt{is} $m_1$\texttt{.+1} \texttt{then}
\vskip0pt\noindent\hskip20pt
    \textit{neodup} (\textit{knuth\_exchange} $m_1$) \texttt{++}
     \textit{ceswap} \texttt{::} \textit{knuth\_jump\_rec} 
    $`2^ m$  $m_1$ (($`2^{m_1}$)\texttt{.-1})
\vskip0pt\noindent\hskip10pt
  \texttt{else} \texttt{[::]}.
\vskip0pt\noindent\hskip0pt
\end{framed}
\vskip0pt\noindent

\section{Batcher's Odd Even Sorter}

The last algorithm we are going to consider is using
both the top-bottom recursion and an even-odd recursion.
For 8 lines, we get.
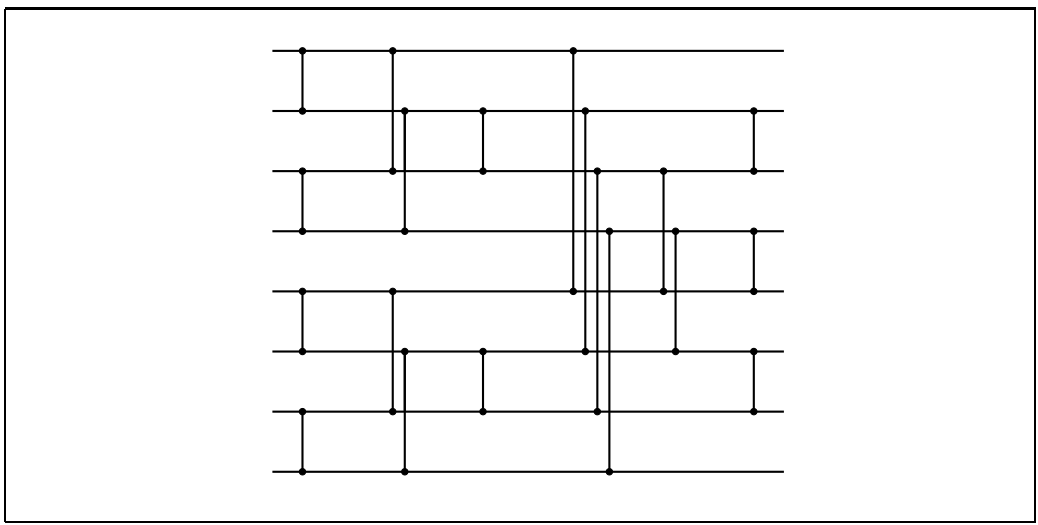
\begin{boxed}
\vspace{5pt}
\vskip5pt
\begin{center}
\begin{tikzpicture}[scale=0.80]
\foreach \a in {1,...,8}
  \draw[thick] (0,\a) -- ++(8.5,0);
\foreach \x in {    
{0.5,1},
{0.5,2},
{0.5,3},
{0.5,4},
{0.5,5},
{0.5,6},
{0.5,7},
{0.5,8},
{2,2},
{2,4},
{2,6},
{2,8},
{2.2,1},
{2.2,3},
{2.2,5},
{2.2,7},
{3.5,2},
{3.5,3},
{3.5,6},
{3.5,7},
{5,4},
{5.2,3},
{5.4,2},
{5.6,1},
{5.6,5},
{5.4,6},
{5.2,7},
{5,8},
{6.5,4},
{6.5,6},
{6.7,3},
{6.7,5},
{8,2},
{8,3},
{8,4},
{8,5},
{8,6},
{8,7}}
  \filldraw (\x) circle (1.5pt);
\draw[thick] (0.5,1) -- (0.5,2);
\draw[thick] (0.5,3) -- (0.5,4);
\draw[thick] (0.5,5) -- (0.5,6);
\draw[thick] (0.5,7) -- (0.5,8);
\draw[thick] (2,2) -- (2,4);
\draw[thick] (2,6) -- (2,8);
\draw[thick] (2.2,1) -- (2.2,3);
\draw[thick] (2.2,5) -- (2.2,7);
\draw[thick] (2.2,2) -- (2.2,3);
\draw[thick] (2.2,6) -- (2.2,7);
\draw[thick] (3.5,2) -- (3.5,3);
\draw[thick] (3.5,6) -- (3.5,7);
\draw[thick] (5,4) -- (5,8);
\draw[thick] (5.2,3) -- (5.2,7);
\draw[thick] (5.4,2) -- (5.4,6);
\draw[thick] (5.6,1) -- (5.6,5);
\draw[thick] (6.5,4) -- (6.5,6);
\draw[thick] (6.7,3) -- (6.7,5);
\draw[thick] (8,2) -- (8,3);
\draw[thick] (8,4) -- (8,5);
\draw[thick] (8,6) -- (8,7);
\end{tikzpicture}
\end{center}
\vskip5pt
\end{boxed}
\noindent
This sorter uses only two connectors. The \textit{cswap} connector is used
in the base case for sorting two lines.
The \textit{codd\_jump} connector with a jump of one is used at the end of the iteration
to get the sorted result when it is sure that the numbers of 
\texttt{false} of the even part exceeds of at most 2 the ones of 
the odd part.
\vskip0pt\noindent
\begin{framed}
\footnotesize
\vskip0pt\noindent\hskip0pt
\texttt{Definition} \textit{batcher\_merge} $m$ : \textit{connector} $m$ \texttt{:=}
 \textit{codd\_jump} 1.
\vskip0pt\noindent\hskip0pt
\texttt{Lemma} \textit{sorted\_batcher\_merge} $m$ ($t$ : ($m + m$)\texttt{.-tuple}
 \texttt{bool}) \texttt{:}
\vskip0pt\noindent\hskip10pt
\textit{noF} (\textit{otake} $t$) $\le$
\texttt{noF} (\textit{etake} $t$) $\le$ (\textit{noF} (\textit{otake} $t$))\texttt{.+2}
 \texttt{->}
\vskip0pt\noindent\hskip10pt
 \texttt{sorted}$_\le$ (\textit{etake} $t$) \texttt{->}
  \texttt{sorted}$_\le$ (\textit{otake} $t$) \texttt{->}
\vskip0pt\noindent\hskip10pt
 \texttt{sorted}$_\le$ (\textit{cfun} \textit{batcher\_merge} $t$).
\end{framed}
\vskip0pt\noindent
In order to sort the odd and even parts, the sorter uses an odd and even
recursion.
\vskip0pt\noindent
\begin{framed}
\footnotesize
\vskip0pt\noindent\hskip0pt
\texttt{Fixpoint} \textit{batcher\_merge\_rec\_aux} $m$ : 
\textit{network} $`2^{m.+1}$ \texttt{:=}
\vskip0pt\noindent\hskip10pt
  \texttt{if} $m$ \texttt{is} $m_1$\texttt{.+1} \texttt{then}
  \texttt{rcons} (\textit{neodup} (\textit{batcher\_merge\_rec\_aux} $m_1$)) 
  \textit{batcher\_merge}
\vskip0pt\noindent\hskip10pt
  \texttt{else} \texttt{[::} \textit{cswap} \texttt{ord0} \texttt{ord\_max}\texttt{]}.
\vskip0pt\noindent\hskip0pt
\texttt{Definition} \textit{batcher\_merge\_rec} $m$ \texttt{:=} 
\vskip0pt\noindent\hskip10pt
  \texttt{if} $m$ \texttt{is} $m_1$\texttt{.+1} \texttt{then}
   \textit{batcher\_merge\_rec\_aux} $m_1$ \texttt{else} \texttt{[::]}.
\end{framed}
\vskip0pt\noindent
The idea is the following. If the top-half and the bottom-half 
are sorted, their respective odd and even part differ at most of one
in the number of \texttt{false}
(the odd part being the smallest).
When taking the odd part and even part of all the lines,
it then differs of at most 2. After sorting them, we are within the conditions of theorem
\textit{sorted\_batcher\_merge}. As having top-half and bottom-half
sorted is preserved by taking the odd or the even part, we get 
the following theorem.
\vskip0pt\noindent
\begin{framed}
\footnotesize
\vskip0pt\noindent\hskip0pt
\texttt{Lemma} \textit{sorted\_nfun\_batcher\_merge\_rec} $m$ 
($t$ : $`2^{m.+1}$\texttt{.-tuple bool}) \texttt{:}
\vskip0pt\noindent\hskip10pt
  \texttt{sorted}$_\le$ (\texttt{take} $`2^{m}$ $t$) \texttt{->}
   \texttt{sorted}$_\le$ (\texttt{drop} $`2^{m}$ $t$) \texttt{->}
\vskip0pt\noindent\hskip10pt
  \texttt{sorted}$_\le$ (\textit{nfun} 
  (\textit{batcher\_merge\_rec\_aux} $m$) $t$).
\end{framed}
\vskip0pt\noindent
We are almost done. We can use top-bottom recursion to fullfill
the conditions of theorem \textit{sorted\_nfun\_batcher\_merge\_rec}.
\vskip0pt\noindent
\begin{framed}
\footnotesize
\vskip0pt\noindent\hskip0pt
\texttt{Fixpoint} \textit{batcher} $m$ : \textit{network} $`2^ m$ \texttt{:=}
\vskip0pt\noindent\hskip10pt
  \texttt{if} $m$ \texttt{is} $m_1$\texttt{.+1}
   \texttt{then} \textit{ndup} (\textit{batcher} $m_1$) \texttt{++}
    \textit{batcher\_merge\_rec} $m_1$\texttt{.+1}
\vskip0pt\noindent\hskip0pt
  \texttt{else} \texttt{[::]}.
\end{framed}
\vskip0pt\noindent
and we get the expected properties.
\vskip0pt\noindent
\begin{framed}
\footnotesize
\vskip0pt\noindent\hskip0pt
\texttt{Lemma} \textit{sorting\_batcher} $m$ : \textit{batcher} m \texttt{is} 
\texttt{sorting}.
\vskip0pt\noindent\hskip0pt
\texttt{Lemma} \textit{size\_batcher} $m$ : \texttt{size} (\textit{batcher} $m$)
 \texttt{=} ($m * m$\texttt{.+1})\texttt{./2}.
\end{framed}
\vskip0pt\noindent
Here is the complete code of the algorithm.
\vskip0pt\noindent
\begin{framed}
\footnotesize
\vskip0pt\noindent\hskip0pt
\texttt{Fixpoint} \textit{batcher\_merge\_rec\_aux} $m$ : 
\textit{network} $`2^{m.+1}$ \texttt{:=}
\vskip0pt\noindent\hskip10pt
  \texttt{if} $m$ \texttt{is} $m_1$\texttt{.+1} \texttt{then}
  \texttt{rcons} (\textit{neodup} (\textit{batcher\_merge\_rec\_aux} $m_1$)) 
  \textit{batcher\_merge}
\vskip0pt\noindent\hskip10pt
  \texttt{else} \texttt{[::} \textit{cswap} \texttt{ord0} \texttt{ord\_max}\texttt{]}.
\vskip0pt\noindent\hskip0pt
\texttt{Definition} \textit{batcher\_merge\_rec} $m$ \texttt{:=} 
\vskip0pt\noindent\hskip10pt
  \texttt{if} $m$ \texttt{is} $m_1$\texttt{.+1} \texttt{then}
   \textit{batcher\_merge\_rec\_aux} $m_1$ \texttt{else} \texttt{[::]}.
\vskip0pt\noindent\hskip0pt
\texttt{Fixpoint} \textit{batcher} $m$ : \textit{network} $`2^ m$ \texttt{:=}
\vskip0pt\noindent\hskip10pt
  \texttt{if} $m$ \texttt{is} $m_1$\texttt{.+1}
   \texttt{then} \textit{ndup} (\textit{batcher} $m_1$) \texttt{++}
    \textit{batcher\_merge\_rec} $m_1$\texttt{.+1}
\vskip0pt\noindent\hskip0pt
\end{framed}
\vskip0pt\noindent

\section{Extension}

A standard extension is to use oriented comparator. Graphically, the orientation
indicates which line gets the maximum of the two lines. This means that, so
far, we have been using comparator with the arrow down.
\vskip5pt
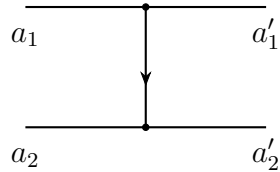
\begin{center}
\begin{tikzpicture}[scale=0.80]
\foreach \a in {1,2}
  \draw[thick] (0,5 - 2 * \a) node[anchor=north] {\vphantom{$a'$}$a_\a$} -- ++(4,0) 
  node[anchor=north] {$a'_\a$};;
\foreach \x in {{2,1},{2,3}}
  \filldraw (\x) circle (1.5pt);
\draw[thick] (2,3) -- (2,1) node[
    sloped,
    pos=0.6,
    allow upside down]{\arrowIn};;
\end{tikzpicture}
\end{center}
\vskip5pt
\noindent
Instead, with the arrow up, the value of the upper line is the maximun of the 
input values, $a'_1 = \max(a_1, a_2)$, and the output value of the lower line 
is the minimum of the two lines, $a'_1 = \max(a_1, a_2)$.
\vskip5pt
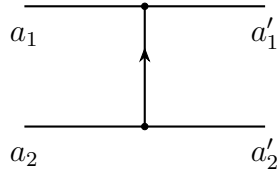
\begin{center}
\begin{tikzpicture}[scale=0.80]
\foreach \a in {1,2}
  \draw[thick] (0,5 - 2 * \a) node[anchor=north] {\vphantom{$a'$}$a_\a$} -- ++(4,0) 
  node[anchor=north] {$a'_\a$};;
\foreach \x in {{2,1},{2,3}}
  \filldraw (\x) circle (1.5pt);
\draw[thick] (2,1) -- (2,3) node[
    sloped,
    pos=0.6,
    allow upside down]{\arrowIn};;

\end{tikzpicture}
\end{center}
\vskip5pt
\noindent
In our formalisation, this means that we need to add an extra component 
that keeps the orientation of the link. This is the field \texttt{cflip}
that associates a boolean to every line. The field \texttt{cflipinv}
ensures that associated lines have identical flip value.
\vskip0pt\noindent
\begin{framed}
\footnotesize
\vskip0pt\noindent\hskip0pt
\texttt{Record} \textit{connector} ($m$ \texttt{:} \textit{nat}) \texttt{:=}
 \textit{connector\_of} \texttt{\{}
\vskip0pt\noindent\hskip21pt
  \textit{clink} \texttt{:} \texttt{\{ffun} \texttt{I\_}$m$ \texttt{->} 
  \texttt{I\_}$m$\texttt{\};}
\vskip0pt\noindent\hskip23pt
  \textit{cflip} \texttt{:} \texttt{\{ffun} \texttt{I\_}$m$ \texttt{->} 
  \texttt{bool}\texttt{\};}
\vskip0pt\noindent\hskip21pt
  \textit{cfinv} \texttt{:} \texttt{[forall} $i$, \textit{clink} 
  (\textit{clink} $i$) \texttt{==} $i$\texttt{];}
\vskip0pt\noindent\hskip10pt
  \textit{cflipinv} \texttt{:} \texttt{[forall} $i$, \textit{cflip} 
  (\textit{clink} $i$) \texttt{==} \textit{cflip} $i$\texttt{]\}.}
\end{framed}
\vskip0pt\noindent
These modifications change the way we define \textit{cfun}
\vskip0pt\noindent
\begin{framed}
\footnotesize
\vskip0pt\noindent\hskip0pt
\texttt{Definition} \textit{cfun} $c$ $t$\texttt{:=}
\vskip0pt\noindent\hskip10pt
    \texttt{[tuple} \texttt{let} \textit{min}\, \texttt{:=} \texttt{min}
     (\texttt{tnth} $t$ $i$) (\texttt{tnth} $t$ (\textit{clink} $c$ $i$)) 
     \texttt{in}
\vskip0pt\noindent\hskip45pt
           \texttt{let} \textit{max} \texttt{:=} \texttt{max} (\texttt{tnth}
          $t$ $i$) (\texttt{tnth} $t$ (\textit{clink} $c$ $i$)) \texttt{in}
\vskip0pt\noindent\hskip45pt
           \texttt{if} $i$ $\le$ \textit{clink} $c$ $i$
           \texttt{then} \texttt{if} \textit{cflip} $c$ $i$ \texttt{then}
          \textit{max} \texttt{else} \textit{min}
\vskip0pt\noindent\hskip45pt
          \texttt{else} \texttt{if} \textit{cflip} $c$ $i$ 
          \texttt{then} \textit{min} \texttt{else} \textit{max} \texttt{|}
           $i$ $<$ $m$\texttt{].}
\end{framed}
\vskip0pt\noindent
The main algorithm that benefits from having this new capability is the bitonic
sorter.
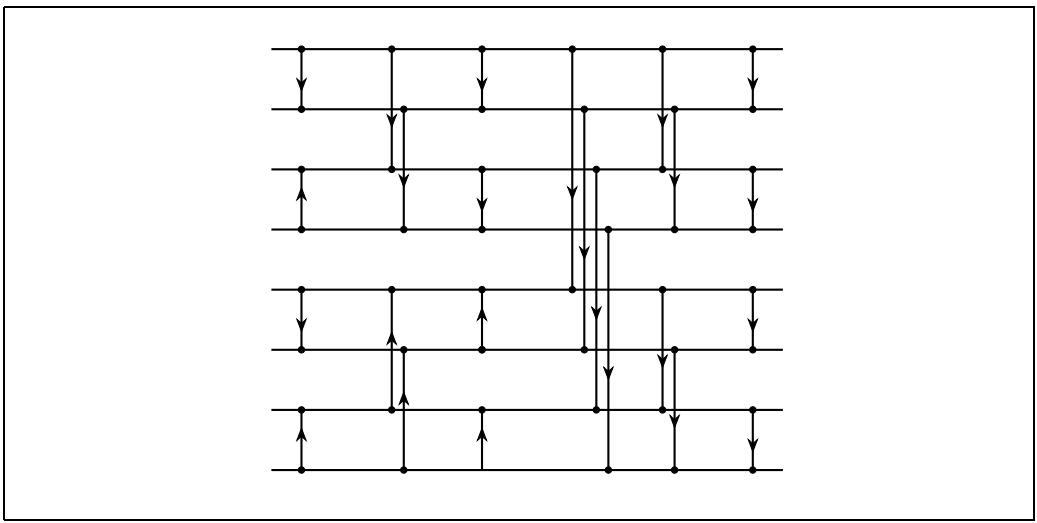
\begin{boxed}
\vspace{5pt}
\vskip5pt
\begin{center}
\begin{tikzpicture}[scale=0.80]
\foreach \a in {1,...,8}
  \draw[thick] (0,\a) -- ++(8.5,0);
\foreach \x in {    
{0.5,1},
{0.5,2},
{0.5,3},
{0.5,4},
{0.5,5},
{0.5,6},
{0.5,7},
{0.5,8},
{2,2},
{2,4},
{2,6},
{2,8},
{2.2,1},
{2.2,3},
{2.2,5},
{2.2,7},
{3.5,3},
{3.5,2},
{3.5,3},
{3.5,4},
{3.5,5},
{3.5,6},
{3.5,7},
{3.5,8},
{5,4},
{5.2,3},
{5.4,2},
{5.6,1},
{5.6,5},
{5.4,6},
{5.2,7},
{5,8},
{6.5,2},
{6.5,4},
{6.5,8},
{6.5,6},
{6.7,1},
{6.7,3},
{6.7,7},
{6.7,5},
{8,1},
{8,2},
{8,3},
{8,4},
{8,5},
{8,6},
{8,7},
{8,8}}
  \filldraw (\x) circle (1.5pt);
\draw[thick] (0.5,1) -- (0.5,2)node[
    sloped,
    pos=0.6,
    allow upside down]{\arrowIn};;
\draw[thick] (0.5,4) -- (0.5,3)node[
    sloped,
    pos=0.6,
    allow upside down]{\arrowIn};;
\draw[thick] (0.5,5) -- (0.5,6)node[
    sloped,
    pos=0.6,
    allow upside down]{\arrowIn};;
\draw[thick] (0.5,8) -- (0.5,7)node[
    sloped,
    pos=0.6,
    allow upside down]{\arrowIn};;
\draw[thick] (2,2) -- (2,4)node[
    sloped,
    pos=0.6,
    allow upside down]{\arrowIn};;
\draw[thick] (2,8) -- (2,6)node[
    sloped,
    pos=0.6,
    allow upside down]{\arrowIn};;
\draw[thick] (2.2,1) -- (2.2,3)node[
    sloped,
    pos=0.6,
    allow upside down]{\arrowIn};;
\draw[thick] (2.2,7) -- (2.2,5)node[
    sloped,
    pos=0.6,
    allow upside down]{\arrowIn};;
\draw[thick] (3.5,1) -- (3.5,2)node[
    sloped,
    pos=0.6,
    allow upside down]{\arrowIn};;
\draw[thick] (3.5,3) -- (3.5,4)node[
    sloped,
    pos=0.6,
    allow upside down]{\arrowIn};;
\draw[thick] (3.5,6) -- (3.5,5)node[
    sloped,
    pos=0.6,
    allow upside down]{\arrowIn};;
\draw[thick] (3.5,8) -- (3.5,7)node[
    sloped,
    pos=0.6,
    allow upside down]{\arrowIn};;
\draw[thick] (5,8) -- (5,4)node[
    sloped,
    pos=0.6,
    allow upside down]{\arrowIn};;
\draw[thick] (5.2,7) -- (5.2,3)node[
    sloped,
    pos=0.6,
    allow upside down]{\arrowIn};;
\draw[thick] (5.4,6) -- (5.4,2)node[
    sloped,
    pos=0.6,
    allow upside down]{\arrowIn};;
\draw[thick] (5.6,5) -- (5.6,1)node[
    sloped,
    pos=0.6,
    allow upside down]{\arrowIn};;
\draw[thick] (6.5,4) -- (6.5,2)node[
    sloped,
    pos=0.6,
    allow upside down]{\arrowIn};;
\draw[thick] (6.5,8) -- (6.5,6)node[
    sloped,
    pos=0.6,
    allow upside down]{\arrowIn};;
\draw[thick] (6.7,3) -- (6.7,1)node[
    sloped,
    pos=0.6,
    allow upside down]{\arrowIn};;
\draw[thick] (6.7,7) -- (6.7,5)node[
    sloped,
    pos=0.6,
    allow upside down]{\arrowIn};;
\draw[thick] (8,2) -- (8,1)node[
    sloped,
    pos=0.6,
    allow upside down]{\arrowIn};;
\draw[thick] (8,4) -- (8,3)node[
    sloped,
    pos=0.6,
    allow upside down]{\arrowIn};;
\draw[thick] (8,6) -- (8,5)node[
    sloped,
    pos=0.6,
    allow upside down]{\arrowIn};;
\draw[thick] (8,8) -- (8,7)node[
    sloped,
    pos=0.6,
    allow upside down]{\arrowIn};;
\end{tikzpicture}
\end{center}
\vskip5pt
\end{boxed}
\noindent
The drawing is more regular since it uses the \textit{half\_cleaner} connector
only.

\vskip0pt\noindent
\begin{framed}
\footnotesize
\vskip0pt\noindent\hskip0pt
\texttt{Lemma} \textit{cflip\_default} $m$ (\textit{clink} \texttt{:}
 \texttt{\{ffun} \texttt{I\_}$m$ \texttt{->} \texttt{I\_}$m$\texttt{\}}) 
 ($b$ \texttt{:} \texttt{bool}) \texttt{:}
\vskip0pt\noindent\hskip10pt
  \texttt{[forall} $i$, \texttt{[ffun} \texttt{=>} $b$\texttt{]}
   (\textit{clink} $i$) \texttt{==} \texttt{[ffun} \texttt{=>} $b$\texttt{]}
    $i$\texttt{].}
\vskip0pt\noindent\hskip0pt
\texttt{Definition} \textit{half\_cleaner} $b$ $m$ \texttt{:=}
\vskip0pt\noindent\hskip10pt
  \textit{connector\_of} (\textit{clink\_half\_cleaner\_proof} $m$) 
  (\textit{cflip\_default} (\textit{clink\_half\_cleaner} $m$) $b$).
\end{framed}
\vskip0pt\noindent
It is now possible to write a version of the bitonic sorter \textit{bfsort}
that uses the flip.

\vskip0pt\noindent
\begin{framed}
\footnotesize
\vskip0pt\noindent\hskip0pt
\texttt{Fixpoint} \textit{half\_cleaner\_rec} $b$ $m$\texttt{:} 
\textit{network} `$2^ m$ \texttt{:=}
\vskip0pt\noindent\hskip10pt
  \texttt{if} $m$ \texttt{is} $m_1$\texttt{.+1} \texttt{then} 
  \textit{half\_cleaner} $b$ `$2^ {m_1}$ \texttt{::}
   \textit{ndup} (\textit{half\_cleaner\_rec} $b$ $m_1$)
\vskip0pt\noindent\hskip10pt
  \texttt{else} \texttt{[::].}
\vskip5pt\noindent\hskip0pt
\texttt{Fixpoint} \textit{bfsort} ($b$ \texttt{:} \texttt{bool}) $m$ \texttt{:}
 \textit{network} `$2^ m$ \texttt{:=}
\vskip0pt\noindent\hskip10pt
  \texttt{if} $m$ \texttt{is} $m_1$\texttt{.+1} \texttt{then} \textit{nmerge}
   (\textit{bfsort} $b$ $m_1$) (\textit{bfsort} (\texttt{\textctilde\textctilde}$b$) $m_1$) 
   \texttt{++}
\vskip0pt\noindent\hskip106pt
                     \textit{half\_cleaner\_rec} $b$ $m_1$\texttt{.+1} 
\vskip0pt\noindent\hskip10pt
  \texttt{else} \texttt{[::].}
\vskip5pt\noindent\hskip0pt
\texttt{Lemma} \textit{size\_bfsort} $b$ $m$ : \texttt{size} 
 (\textit{bfsort} $b$ $m$) \texttt{=} ($m * m$\texttt{.+1})\texttt{./2}.
\vskip0pt\noindent\hskip0pt
\texttt{Lemma} \textit{sorting\_bfsort} $m$ : \textit{bfsort} \texttt{false}
 $m$ \texttt{is} \textit{sorting}.
\end{framed}
\vskip0pt\noindent

\section{Conclusion}

In this paper, we have shown how to formalise different sorting 
algorithms for networks. We have been following mostly what is presented
in chapter 28 of \cite{cormen}. Another source of inspiration was~\cite{bitonic}.
We have been using intensively the zero-one principle.
Most of the proof are done manipulating booleans. It looks a bit 
like magic. The formalisation is available at

\url{https://github.com/thery/mathcomp-extra}

\noindent
It consists of 5 files. The file \texttt{more\_tuple} contains
some addition to the Mathematical Library. It is  
1000-line long. The file \texttt{nsort} contains
the definition of network and some basic connectors. 
It is 700-line long.
The file \texttt{bitonic} deals with the bitonic sorter. 
It is 500-line long.
The file \texttt{bjsort} deals with the exchange sorter. 
It is 200-line long.
The file \texttt{batcher} deals with the exchange sorter. 
It is 200-line long.
 
From the specification point of view, we believe 
that having explicit networks  and using 
dependent types for this gives us a very concise presentation of the 
algorithms. All the usual index manipulations are hidden
inside the \textit{ndup} and \textit{neodup} building blocks.
From the proving point of view, the difficult part in the bitonic sort is
proving the specification of the half-cleaner. From the other 
sorters, the only delicate thing is the manipulation of \textit{codd\_jump}
connectors. The introduction of the function \textit{noF} makes 
the specification and proof easier.

\bibliographystyle{plain}
\bibliography{Note}

\appendix

\section{Basic \label{basic}}

$$
\begin{array}{|r@{\quad}|@{\quad}l|}
\hline
& \cr
\hline
\texttt{$x$\,\,\texttt{=}\,\,$y$}
    & \textrm{propositional equality between $x$ and $y$}\cr                           
\texttt{$x$\,\,\texttt{==}\,\,$y$}
    & \textrm{boolean equality between $x$ and $y$ that must belong to an eqType}\cr                           
\texttt{reflect}\,\, P \,\, b &
      \textrm{equivalence between the propositions $P$ and ($b$ = \texttt{true})}
      \cr
n\texttt{.+1}
    & \textrm{add one to the natural number $n$}\cr                           
n\texttt{.*2}
    & \textrm{double the natural number $n$}\cr                           
n\texttt{./2}
    & \textrm{half the natural number $n$}\cr                           
\texttt{uphalf } n
    & \textrm{half the natural number $n + 1$}\cr                           
\texttt{odd}\,\,n
    & \textrm{\texttt{true} if $n$ is odd, \texttt{false} otherwise}\cr                           
\texttt{(}l\texttt{,}\,\, r{)}
    & \textrm{the pair composed of $r$ and $l$}\cr                           
p\texttt{.1}
    & \textrm{the first component of the pair $p$}\cr                           
p\texttt{.2}
    & \textrm{the second component of the pair $p$}\cr                         
\texttt{[qualify}\,\,x\,\,\texttt{|}\,\,P\texttt{]}
    & \textrm{if \texttt{$A$ := [qualify}$\,\,x\,\,$\texttt{|}$\,\,P$\texttt{]},
      $x$ \texttt{is} $A$ is equivalent to $P$ }\cr
\hline
\end{array}
$$
\section{Fintype \label{fintype}}

$$
\begin{array}{|r@{\quad}|@{\quad}l|}
\hline

\texttt{[forall}\,\,x\texttt{,}\,\, P\texttt{]}
    & \textrm{$P$ (in which $x$ can appear) is true for all values of $x$}\cr
    & \textrm{$x$ must range over a finType}\cr                           
\texttt{I\_}n & \textrm{the finite subType of integers}\ 
 $i < n$ \cr
  \texttt{ord0} & \textrm{ the $i$ : \texttt{I\_}$n$\texttt{.+1} with value $0$}\cr
  
  \texttt{ord\_max} & \textrm{ the $i$ : \texttt{I\_}$n$\texttt{.+1} with value $n$}\cr
    \texttt{inZp} & \textrm{the natural projection from nat into the integers 
    mod $p$, represented}\cr
    &\textrm{ as {\textquotesingle}\texttt{I\_}$p$. 
    Here $p$ is implicit, but must be of the form $n$\texttt{.+1}}\cr  
   \texttt{rev\_ord}\,\, i & \textrm{the complement to $n$\texttt{.-1}
                                      of $i$ : \texttt{I\_}$n$} \cr
   \texttt{lshift}\,\, n\,\, j &
   \textrm{the $i$ : {\textquotesingle}\texttt{I\_}($m$+ $n$) 
   with value $j$ : {\textquotesingle}\texttt{I\_}$m$}\cr
   \texttt{rshift}\,\, m\,\, k &
   \textrm{the $i$ : {\textquotesingle}\texttt{I\_}($m$+ $n$) 
   with value $m + k$, $k$ : {\textquotesingle}\texttt{I\_}$n$}\cr
\texttt{split}\,\, i &
 \textrm{$i$ has type {\textquotesingle}\texttt{I\_}($m$ + $n$)}\cr
 & \textrm{it returns \texttt{inl} $j$ \textrm{when there exists $j$ such}}\cr 
 & \hskip115pt \textrm{that $i$ = \texttt{lshift} $n$ $j$}\cr
 & \textrm{it returns \texttt{inr} $k$ when there exists $k$ such}\cr 
 & \hskip115pt \textrm{that $i$ = \texttt{rshift} $m$ $k$}\cr
\texttt{\{ffun}\,\,A\,\, \limp\,\, B\texttt{\}} &
\textrm{type for functions with a finite domain ($A$ should be a finType)}\cr
\texttt{[ffun}\,\,x\,\, \texttt{=>}\,\, E\texttt{]} &
\textrm{definition of a function with a finite domain ($x$ may appear in $E$)}\cr
 \hline
\end{array}
$$

\noindent
\section{Sequences \label{seq}}
$$
\begin{array}{|r@{\quad}|@{\quad}l|}
\hline
& \cr
\hline
  \texttt{[::]} &  \textrm{the empty sequence} \cr
  x\,\,\texttt{::}\,\,s & \textrm{the sequence starting with $x$ followed by $s$}\cr
  \texttt{rcons}\,\,s\,\, x & \textrm{the sequence starting with $s$ and ended by $x$}\cr
\texttt{[seq}\,\,E\texttt{|}\,\, i\,\,\texttt{<-}\,\, l\texttt{]} &
\textrm{the sequence with general term\ }
 E \textrm{ ($i$ in $l$ and bound in  $E$ )} \cr
   \texttt{size}\,\, s & \textrm{the number of items (length) in $s$} \cr
  \texttt{count}\,\, P\,\, s & 
  \textrm{the number of items of $s$ for which $P$ holds}\cr 
  \texttt{nseq}\,\, n\,\, x & \textrm{a sequence of $n$ $x$'s}    \cr
  \texttt{head}\,\, x_0\,\, s & \textrm{the head (zero'th item) of $s$ if $s$ 
                        is non-empty, else $x_0$}\cr
  \texttt{behead}\,\, s &  \textrm{$s$ minus its head} \cr
  \texttt{last}\,\, x\,\, s & \textrm{the last element of $x$ 
  \texttt{::} $s$ (which is non-empty)} \cr
  \texttt{belast}\,\, x\,\, s & \textrm{$x$ \texttt{::} $s$ minus its last item}
  \cr
   s_1\,\, \texttt{++}\,\, s_2 &  \textrm{the concatenation of $s_1$ and $s_2$}\cr
   \texttt{take}\,\, n\,\, s & \textrm{the sequence containing only the first 
      $n$ items of $s$}\cr
    &   \textrm{(or all of $s$ if \textrm{size} $s \le n$)} \cr
    \texttt{drop}\,\, n\,\, s & 
    \textrm{$s$ minus its first $n$ items (\texttt{[::]} if \texttt{size} $s \le n$)
         } \cr
   \texttt{rot}\,\,n\,\, s & \textrm{$s$ rotated left $n$ times (or $s$ if 
   \texttt{size} $s$ $\le$ $n$)}\cr
  \texttt{zip}\,\, s\,\, t &  \textrm{itemwise pairing of $s$ and $t$
     (dropping any extra items)} \cr
  & \texttt{[::($x_1$, $y_1$); $\dots$; ($x_{mn}$, $y_{mn}$)]}
   \textrm{with $mn = \texttt{min}\,\, n\,\, m$}\cr
  \texttt{foldl\ }f\ a\ s & 
  \textrm{the left fold of $s$ by $f$,
   i.e. $f$ ($f$ $\dots$ ($f$ $a$ $x_1$) $\dots$ $x_{n-1}$) $x_n$
  }\cr 
 \texttt{perm\_eq $s_1$ $s_2$} &
 \textrm{$s_2$ is a permutation of $s_1$, i.e., $s_1$ and $s_2$ have the
                     items} \cr & 
  \textrm{(with the same repetitions), but possibly in a    
                    different order}
\cr
 \texttt{sorted$_e$ $s$} & 
 \textrm{$s$ is an $e$-sorted sequence: either $s$ = \texttt{[::]},
 $s$ = \texttt{[::$x$]},} \cr
 &  \textrm{or $s$ = \texttt{$x$\,::\,$y$\,::\,$s_1$} with 
$e$ $x$ $y$ and ($y$ \texttt{::} $s_1$) is $e$-sorted}
\cr
\hline
\end{array}
$$

\section{Tuple \label{tuple}}

$$
\begin{array}{|r@{\quad}|@{\quad}l|}
\hline
& \cr
\hline
  n\texttt{.-tuple\ } T &
  \textrm{the type of n-tuples of elements of type}\ T    \cr
\texttt{[tuple\ }E\texttt{\ |\ } i < n\texttt{]} &
\textrm{the\ } n\texttt{.-tuple} \textrm{\ with general term\ }
 E \cr 
&  \textrm{\ (} i \texttt{\ :\ } \texttt{I\_}n 
\textrm{\ is bound in\ } E \textrm{)} \cr
\textrm{tnth\ } t\ i & \textrm{the $i$'th component of $t$, where $i$ :
\texttt{I\_}$n$} \cr 

\hline
\end{array}
$$

\end{document}